\documentclass[aps,pre,groupedaddress,amssymb,amsmath,pra,floatfix,showpacs]{revtex4}
\usepackage{epsfig}

\begin{document}
\title{Quantum Stabilization by Coherent Population of Scarred Eigenstates:
Quantum Suppression of Transport Classically Initiated by Separatrix Band Entry}
\author{Luca C. Perotti}
\affiliation{Department of Physics, Texas Southern University, Houston, Texas 77004 USA}
\date{\today}
\begin{abstract}

Comparisons of experimental data with numerical predictions of a
classical model indicate that an excited hydrogen atom in a pulsed
microwave electric field exhibits a nonclassical increase of
stability over a relatively wide range of frequencies. I show here
that this is due to selective population of long-lived ``scarred"
states that are associated with the chaotic separatrix band
surrounding the principal classical resonance zone in phase space.
A quantum explanation is given in terms of adiabatic evolution of
Floquet states and the destabilizing effect of two-level quantum
resonances is investigated. The role of neighbouring classical
resonance zones in defining the frequency range of stabilization is
revealed both by quasienergy curves and by Husimi functions for the
instantaneous quantum states.

\end{abstract}
\maketitle


\section{Introduction}

Since early in the development of quantum mechanics, it has been
apparent the importance of coherent wavepackets as ``semiclassical"
objects following a classical orbit in phase space
\cite{ref1,ref2}. In most systems this semiclassical behaviour is
relatively short lived, as the packets have to be built as a
superposition of eigenstates of the system \cite{ref3} and
dephasing soon breaks them up. ``Revivals" (reforming of the packet
after some time) are possible in bound systems, due to the
quasi-periodicity of the quantum evolution, a consequence of the
discreteness of the spectrum. Because of that ``quasi" though, most
of the times  these revivals are only partial. The recent discovery
of eigenstates behaving as (nondispersive) wavepackets themselves
\cite{ref4,ref5,ref6} is therefore of great interest and has spurred studies both theoretical \cite{nond} and experimental \cite{nond1}.

One semiclassical dynamical system accessible to experimental study
where these particular ``eigenstates" (actually very long lived
resonances) have been found is the highly excited hydrogen atom
exposed to a microwave electric field. An attractive characteristic
of this system which makes it a convenient system for study is that
if the atom is prepared in a static electric field collinear with
the microwave field, and an extreme Stark state is excited, a
one-dimensional approximation can be used, thus greatly simplifying
the treatment.

The extended classical phase space $(\theta,I,t)$ of the
one-dimensional hydrogen atom in a microwave electric field is
approximately divided into a region of regular motion barely
affected by the microwave field (field- modulated atom), a region
dominated by resonance zones (nested vortex tubes surrounded by
local chaos), and a globally chaotic region. In particular, when
the ratio $\omega'_0
\equiv n_0^3 \omega/(1-3n_0^4F_S) $ of the microwave frequency $\omega$ to the
initial Kepler orbit frequency $(1-3n_0^4F_S) /n_0^3$ (corrected
for the static field $F_S$) is near unity, resonant  classical
motion dominates; we call this region of phase space the
``principal primary resonance zone". Stroboscopic surfaces of
section (Poincar\'{e} maps) reflect these zones in the full phase
space as characteristic zones in the action-angle  $(\theta,I)$
subspaces (``sectioned" phase spaces); in any such subspace a
resonance zone appears as a chain of stable regions (islands)
centered on stable periodic orbits and surrounded by a locally
chaotic region containing unstable periodic orbits. The vortex
tubes produce the resonance island zones and the local chaos
produces the zones of irregular motion that have been called
``separatrix" zones \cite{ref7}: just like the separatrix in
integrable problems, a chaotic layer -that grows with increasing
microwave field strength- separates regions corresponding to
different kinds of regular motion. The same type of structure is
seen repeating itself within the primary resonance zones
themselves, resulting in secondary resonance zones. The details of
this picture change with the strength of the microwave field.

These classical zones in phase space are reflected in the character
of the Floquet eigenstates (FE) \cite{ref8}: regular FE are
modulated atom FE or resonance island FE, supported by the nested
vortex tubes of a resonance island chain. Noticeable among
irregular FE are the separatrix FE, possibly ``scarred" \cite{ref9}
by a high electron probability density in phase space along the
unstable periodic orbits of the classical separatrices that support
them  \cite{ref10,ref4}. At intermediate values of microwave field
strength, localization along the unstable periodic orbits is only
partial and the FE extend over a bounded region of phase space
classically occupied by chaotic trajectories \cite{ref8,ref4}. Ref.
\cite{ref4} shows that a field modulated FE is still similar to a
free atom state: its configuration space wavefunction is almost
stationary, there being no evident sign of significant
oscillations. The FE supported by the center of the principal
primary resonance zone is instead a {\bf nondispersive packet}
oscillating at the frequency of the microwave field, that is the
frequency of the classical {\bf stable} periodic orbit around which
the state is localized  \cite{ref4}. The ``packet" is most of the
time quite smooth, apart from some self interference when it is
close to the nucleus, (the case of the surfaces of section in ref.
\cite{ref11}, where the island is strongly deformed into a heart
shape). This smoothness and almost Gaussian shape of the state is
connected to the high field behaviour of the quasienergy levels of
the principal primary quantum nonlinear resonance. In the high
field limit they have a linear dependence on the ``resonance
quantum number" $k$ (see ref.  \cite{ref17} and ref. therein), that
is: they are equally spaced  \cite{ref13}. The system is therefore
at least locally harmonic and the $k = 0$ quasienergy state
supported by the center of the resonance is essentially the
(Gaussian-shaped) ground state of a harmonic oscillator that, in
the case of the principal primary resonance (the only case
investigated up to now), is ``forced" by the microwave field to
oscillate at the frequency  \cite{ref4} of the microwave itself
\cite{ref14}. Ref.  \cite{ref6} attributes these forced
oscillations to ``a nonlinear coupling between the atom and the
driving field that locks the electronic motion to the driving
frequency" ; this phenomenon is known as {\bf autophasing of the
nonlinear oscillations}  \cite{ref15}.

Finally a separatrix FE is a rather irregular {\bf nondispersive
packet} oscillating with the same period of the microwave, that is:
following the classical {\bf unstable} periodic orbit around which
the state is localized  \cite{ref4}. The irregularity or
``contamination" of the state is attributed by Ref.  \cite{ref4} to
the flow of probability through the separatrix zone between the
inside and the outside of the resonance. To view this quantum flow
in phase space it is necessary to resort to some representation in
phase space of the wavefunctions; Husimi functions have the
advantage on the more commonly used Wigner functions of being
positive defined. Husimi functions of states with support in the
separatrix region between the separatrix state and the primary
island sometimes display branches of probability extending to
higher actions and a superposition of any of these Husimi functions
over the separatrix Husimi function shows that these branches seep
through the probability valleys that lay between the peaks of the
separatrix Husimi function  \cite{ref4}. These valleys are
therefore a suspected quantum equivalent of the separatrix
``turnstiles" described in ref.  \cite{ref16} responsible for the
classical flow of probability out of resonance zone.

The importance of these ``scarred" separatrix wavefunctions is that with increasing microwave field
strength their lifetime decreases much more slowly than that of the states whose support in phase space is
within the resonance zone and for sufficiently strong fields becomes comparable to that of the (``Gaussian")
state at the center of the resonance zone  \cite{ref4}. While the stability of the state at the center of the
resonance zone brings about an agreement between quantum and classical behaviour (``classical"
stabilization), the comparable stability of the ``scarred" wavefunction in a region of chaotic classical
behaviour causes a sharp divergence between quantum and classical behaviour (``quantum" stabilization).

The present paper deals with the possibility of selective
population, starting from a highly excited hydrogen atom, of one
such ``separatrix" state at the peak of a partially ionizing short
pulse of microwave electric field. I show that, taking care to
avoid some critical cases, this is usually possible for the range
of parameters of the laboratory experiments \cite{ref17}. Previous
studies dealing with this form of quantum stabilization
\cite{ref11,ref18,ref19,ref20} concentrated on the sharp $\omega'_0
\simeq 0.69$ maximum in the peak microwave field strength
$F_0(10\%)= F(10\%)n_0^4$ (rescaled to the average Coulomb field on
the Kepler orbit $1/ n_0^4$) necessary for $10\%$ ionization
probability (see Fig. \ref{fig1}); I instead claim that it  is the
divergence between classical simulations (circles) and experimental
results (triangles) over the whole region  $\omega'_0
\in  (0.69, 0.85)$ that is due to the selective population of
$\omega'_0 =1/1$ separatrix states \cite{ref40}. The $\omega'_0
\simeq 0.69$ peak itself only marks the "crossover" point where we pass
from the $\omega'_0 =1/1$ resonance zone to the $\omega'_0 =2/3$
one. To the right of the $\omega'_0 \simeq 0.69$ peak, the
experimental $F_0(10\%)$ curve approximately corresponds to the
initially populated quantum state evolving  (at the peak of the
pulse) into the $\omega'_0 =1/1$ separatrix state (a quantum
adiabatic process). To the left of that same peak the initially
populated quantum state is also the state at the center of the
$\omega'_0 =2/3$ resonance zone. Not only adiabatic population of a
single state at the center of a resonance zone might require a much
longer switch-on time than adiabatic population of a state at the
separatrix \cite{ref21,ref19}, but (more important, as we shall
see) for such a small resonance zone as the $\omega'_0 =2/3$ one,
the state at the center of the resonance zone is less stable than
the separatrix state.

The paper is organized as follows: in section II, I present the
system and the numerical techniques I use to investigate it; the
results of my investigation are presented and interpreted in
section III. Section IV finally sums up my findings.

\section{THE SYSTEM AND NUMERICAL METHODS}

The system I investigate is a 1D  model for a stretched highly excited hydrogen atom in collinear static
and monochromatic microwave electric fields \cite{ref17}. In atomic units the Hamiltonian reads
\begin{eqnarray}
H = {\frac{p^2}  2} - {\frac 1 z} +z[A(t)F^{max} sin{( \omega t + \phi_0)}
- F_S],\nonumber\\
z \geq 0,
\label{eq1}
\end{eqnarray}
where $A(t)$ is the envelope of the microwave pulse. To simulate
the experimental situation described in ref. \cite{ref23}, I have
chosen
\begin{eqnarray}
A(t) =  \sin {\left({\frac{\pi t}{T_p}}\right)}\nonumber
\end{eqnarray}
where $T_p$ is the length of the microwave pulse.

The Hamiltonian \ref{eq1} has the following important scaling
property: let $c$ be an arbitrary constant, introducing the scaled
variables $p'=cp$,  $z'=z/c^2$, $t'=tc^3$, $H' = c^2H$ and the
scaled parameters $F_0^{max}= c^4F^{max}$ (peak microwave field
strength), $F_{S0}= c^4F_S$ (static field) and $\omega_0 =
c^3\omega$ (microwave frequency), we again have
\begin{eqnarray}
H' = {\frac{{p'}^2} 2} - {\frac 1 z'} +z[ \sin {\left({\frac{\pi t}{T_p}}\right)}F_0^{max} sin{( \omega_0 t' +
\phi_0)}
- F_{S0}], \nonumber
\end{eqnarray}
and Hamilton's equations are invariant in form. In this paper I
present my findings in terms of classically scaled parameters,
choosing the constant $c$ to be the quantum number $n_0$ in which
the system is initially prepared. To compensate for the Stark shift
of the atomic frequencies I present my data in terms of the first
order Stark corrected scaled frequency $\omega'_0\equiv n_0^3
\omega_0/(1-3n_0^4F_S) $ \cite{ref24}.

\subsection{CLASSICAL METHODS}

I numerically solved Hamilton's equations of motion in the free-atom action angle variables $(I,\theta)$,
valid when the electron's energy is negative:
\begin{eqnarray}
z = 2I^2 \sin^2 (\xi /2),\nonumber\\ p = (1/I) \cot{(\xi
/2)},\nonumber
\end{eqnarray}
where the eccentric anomaly angle $\xi (\theta)$ is defined by
$\theta \equiv \xi - \sin{\xi}$. To avoid equations of motion
containing terms that diverge as $z$ approaches zero, a dummy time
$\eta$ was introduced, defined by the equation $dt
\equiv (1 -\cos{\xi})d \eta$ \cite{ref26}. The true time $t$ increases monotonically with
$\eta$. The integration was performed using a fixed step, fourth
order Runge-Kutta routine \cite{ref17}. The ensemble of initial
conditions $(\theta_0,I_0)$ was chosen so that, to first order in
Fs, the (classical) electron energy $E(0)$ be equal to the value of
the energy of the experimentally prepared initial quantum energy
eigenstate
\begin{equation}
-{\frac 1 {2n_0^2}} - {\frac 3 2}F_S n_0^2 = E(0) = -{\frac 1 {2I_0^2}} - F_SI_0^2(1 - \cos{\xi_0}).
\label{eq2}
\end{equation}
Given a value of $\theta_0(\xi_0)$, this equation determines a
corresponding value of $I_0$.

Classical values for the ``ionization" probability $P_I$ at the end
of the pulse were averaged over uniform distributions of the
initial angle $\theta_0$ and microwave field phase $\phi_0$ as
already described in ref. \cite{ref17}. The "ionization"
probability was the sum of two contributions. One was from
trajectories that were terminated at some time during the pulse
where the instantaneous value of the energy $E$ exceeded the value
$-2\sqrt{F_S}$ for rapid ionization in the static field alone. The
other contribution was from trajectories whose final value of $E$
corresponded to energies of quantum energy eigenstates, see
equation (\ref{eq2}) above, with quantum numbers outside the
interval $[50,90]$ of detection of the ``survived" atoms
\cite{ref17}. Stroboscopic surfaces of section (Poincar\'{e}  maps)
in $(\theta,I)$ space were also computed for $A(t)F^{max}$ equal to
various constants $F$. They revealed the (long-time) structures
existing in phase space for the various values of $F$ that were
traversed during the microwave pulse. The comparison of
instantaneous ensemble distributions in phase space with these
surfaces of section, for various times $t$ during the pulse has
proven a useful tool for the understanding of the evolution of the
ensemble itself \cite{ref23,ref17}.

\subsection{ QUANTUM METHODS}

The time evolution of the quantum system was evaluated by numerical
integration of the Schrodinger equation with $H$ given by
eq.\ref{eq1} on a finite subset of the (bound) free atom basis
$\psi_n(z)$ chosen large enough so that the probability reflected
at the boundaries was small \cite{ref26}. Given an initial state
$\psi(z,0) = \Sigma_n C_n(0) \psi_n(z)$, the equation for the
evolution of the expansion coefficients $C_n(t)$ is
\begin{eqnarray}
i {\frac {dC_n(t)} {dt}} = E_n C_n(t) + {\mathcal F}(t) \Sigma_m
Z_{n,m} C_m(t), \nonumber\\ {\mathcal F}(t) = (F(t) sin(\omega t +
\phi_0) - Fs),
\label{eq3}
\end{eqnarray}
where $Z_{n,m}$ is the matrix element of the operator $z$ between
the states $n$ and $m$. I approximated ${\mathcal F} (t)$ with the
function ${\mathcal F'}(t) = {\mathcal F}(t)\Delta t \Sigma_k
\delta(t - k\Delta t)$ that tends to ${\mathcal F} (t)$ for $\Delta
t \to 0$ (in the sense that the integral of their difference over
an arbitrary time interval goes to zero as $\Delta t$)
\cite{ref26}. I took into account the loss of probability induced
by the static field by the substitution in equation (3) of $E_n$
with $(E_n-i\Gamma_n/2)$, where the decay factors $\Gamma_n$ are
given by the ($3D$) formula from ref. \cite{ref27}. The above
procedures involved a number of  unphysical parameters that had to
be carefully chosen as not to falsify the results of the
integration. In particular, truncating the basis puts us in a
``fuzzy box"; in absence of microwaves the potential is therefore a
kind of ``double well" where not only the width but also the depth
of the second well varies with the number of levels considered. I
discussed my choices in ref. \cite{ref17}.

\subsubsection{Quasienergy curves}

Like in the classical case, the dynamics at constant microwave
amplitude can help understanding the pulsed dynamics. If $A(t)$ is
a constant, Schrodinger's equation is a differential equation with
time-periodic coefficients. Floquet's theory is therefore
applicable \cite{ref28} and tells us that the equation has
solutions in the form
\begin{eqnarray}
\psi_i (t) = \Phi_i (t) e^{-i\varepsilon_i t/ \hbar}, \hspace{.4in} \Phi_i (t + T)= \Phi_i (t)
\nonumber
\end{eqnarray}
where $T$ is the period of the microwave, the constants
$\varepsilon_i$ take the name of quasienergies and the functions
$\psi_i (t)$ are called quasienergy (or Floquet) states. For
periodic systems quasienergies take the place of energies in the
description of the dynamical properties of the system. As
quasienergies shifted by $2\pi \hbar /T$ with respect to each other
correspond to the same physical state $\psi_i (t)$, it is
sufficient to restrict ourselves to the energy interval $[0, 2\pi
\hbar /T)$ (first Brillouin zone) to have all the levels. To
calculate the quasienergies I resorted to the time evolution
operators $G(\phi)$ over one period $T$ of the microwave (Floquet
operators), parametrized by the phase $\phi= (\omega t_0 + \phi_0)$
of the field at the beginning of the period. While the
eigenfunctions $\Phi_i (\phi)$ of each member of the family are
different and represent the different spatial structures of the
states at different times during the microwave period, the family
shares the same eigenvalues $G_i$, connected to the quasienergies
by the relationship $G_i = e^{-i\varepsilon_i T/\hbar}$. To obtain
the quasienergies I therefore numerically calculated a one period
evolution operator and diagonalized it; the above equation then
gave us the quasienergies on the circle $[0, 2\pi
\hbar /T)$.

Plots of the quasienergies as functions of the microwave field
strength $F$, called {\bf quasienergy curves} (see Fig.
\ref{fig2}), help us connect the characteristics of the system at
different values of the microwave field strength. The quasienergies
change with $F$ and undergo a sequence of avoided crossings of
widely different widths. Since in the $1D$ hydrogen atom all the
quasienergies fall into the same symmetry class \cite{ref29}, the
Von Neumann-Wigner theorem \cite{ref30} tells us that they all
repel each other; this implies the absence of real level crossings.
As in our simulations of the pulsed system, we calculate the
Floquet operator on a truncated basis and some caution has to be
exerted to avoid spurious results. A study of the quasienergy
curves themselves can help us in deciding what parts of what curves
to believe \cite{ref31,ref53,ref17}.

\subsubsection{Husimi function}

When studying semiclassical systems, one wishes to understand to
what extent phase space classical stuctures can guide the quantum
evolution. To this end it is useful to have some representation of
wavefunctions in phase space. One of the most widely used such
representation is the Husimi function \cite{ref32}. Given a quantum
function $\psi(t)$, its Husimi function at a point
$\{\left<{p}\right>,\left<{q}\right>\}$ in phase space is the
projection (in atomic units)
$\rho_H(\left<{p}\right>,\left<{q}\right>,t) = | \left<{
\phi_{\left<{p}\right>\left<{q}\right>} | \psi(t) }\right> |^2$ of $\psi(t)$
on a minimum uncertainty packet centered on that point
in phase space and called the ``coarse graining function". In
action-angle space, due to the periodicity in the angle variable,
the standard (position-momentum space) choice of coarse graining
function [32] is not possible. Following ref. [11] I therefore
take
\begin{eqnarray}
\phi_{\left<{I}\right>\left<{\theta}\right>}=\Sigma_{n=0}^{\infty}\left[{{\frac {\alpha^{(\alpha n+1)}}
{2\pi \Gamma(\alpha n+1)}} \left<{I}\right>^{\alpha n} e^{-\alpha
\left<{I}\right>}}\right]^{1/2} e^{i2\pi \left<{\theta}\right> n}\psi_n.\nonumber
\end{eqnarray}
Tests with different values of $\alpha < n_0$ have given (both for
ref. \cite{ref11} and for me) Husimi functions having essentially
the same shape. The comparison of the Husimi function of the
wavefunction at various times of the microwave pulse with the
classical ensemble at the same times and with the classical
surfaces of section at the same values of $\omega$, $F$ and
$\phi_0$ can often  be quite instructive.

\section{Results}

\subsection{QUANTUM ADIABATIC BEHAVIOUR AT THE SEPARATRIX BAND}

The behaviour of a grouping of strongly interacting levels
(repelling each other) at, or close to zero microwave field, is
particularly evident in Fig. \ref{fig2}. I have marked them as
darker lines. The (zero microwave field) energy of each of these
levels is close to a quantum resonance $\Delta E \equiv  E_{n+r} -
E_n = s \omega \hbar$ with any of the other levels of the same
grouping; the ratio $s/r$ being equal to $1$ for any two levels of
the grouping. This grouping represents on the $(E,F)$ plane what
ref. \cite{ref33} calls a nonlinear quantum resonance : a finite
(because of level anharmonicity) number of quantum states whose
interaction is due to the existence of a (primary) classical
resonance zone, in this case the $\omega'_0 =1/1$ resonance zone.
The levels initially curve downwards, reach a minimum and then
start growing at a fast rate; the inflection point of a curve
indicates where the character of the state changes from field
modulated free atom to resonance state \cite{ref17}; the transition
is drastic (see appendix B) and is characterized by a change in
semiclassical WKB quantization \cite{ref17}. At lower microwave
field strengths the quantum number is the free atom one, $n$; while
for higher fields the quantum number is the ``resonance" one,
$k=0,1,2 ...$ The $k=0$ state at the center of the quantum linear
resonance being the one with the fastest growing quasienergy.
Therefore the inflection point is where we can find a separatrix
state. If, for the pulsed system we are considering, we want to
selectively populate such a state at the peak of the pulse, we must
choose an initial state such that the quasienergy curve originating
from it has its inflection point close to the peak of the pulse
itself.

As an example to be viewed through Figure \ref{fig2}, let us
consider a Hydrogen atom prepared in an extreme Stark state with
principal quantum number $n_0= 65$ in a static field $F_S = 8
\hspace{.1in} V/cm$, and subject it to a microwave pulse with $\omega
= 18.00 hspace{.1in} GHz$, and $F^{max}= 4.63 \hspace{.1in}V/cm$, lasting $T =
140$ microwave periods; rescaled, the parameters of the microwave
and static fields are $\omega'_0 =0.8196$, $F_0^{max} = 0.01607$
and $F_{S0}=0.02777$. The principal primary resonance is centered
at $n_r = 69$ and according to the equation at the top of page 2185
of ref. \cite{ref19}, at the peak of the pulse the resonance
contains
\begin{eqnarray}
N = {\frac {8n_r} {\pi}}\left({\frac {0.325F_r^{max}} {3(1+F_{rs})}}\right)^{1/2} = 8\nonumber
\end{eqnarray}
quantum states (see appendix A for a discussion of the small
difference between the present formula and the one in ref.
\cite{ref19}). A look at the zero microwave field ``quasienergies"
moreover shows that these $8$ states are the states $n=65$ to
$n=72$. A classical ensemble simulating at $t=0$ the quantum Stark
state $n_0 = 65$ will therefore enter the resonance zone close to
the peak of the pulse, as can be seen in Fig. \ref{fig3} which
shows snapshots of the classical ensemble near the peak and at the
end of the pulse: the ensemble enters the primary resonance zone
after about $55$ periods of the microwave, when the field strength
is about $94\%$ of peak. The full line in Fig. \ref{fig4} shows the
final distribution in action that is, as expected \cite{ref34},
double peaked in action with one peak centered on $n_0= 65$ and the
other one centered on its ``symmetric" action ($n_s \simeq 73$)
with respect to the center of the island ($n_r=69$). It is also
evident, both from Fig. \ref{fig4} and from the last snapshot of
Fig. \ref{fig3}, that the two peaks are rather wide and of quite
different shapes. Even if there is no noticeable spread in action
at the crossing itself, the width of the peaks comes as no surprise
if we consider the wide chaotic separatrix region over which the
ensemble is spread at the peak of the pulse (see the snapshot at
$t=70$ microwave periods in Fig. \ref{fig3}). The asymmetry in
shape of the two peaks takes the form of an approximately
exponential fall of the population distribution for $n > n_0 = 65$,
superimposed over the double peak structure (see Fig. \ref{fig4}).
This shape suggests a coupling of the chaotic band to the global
chaos above as the cause of the asymmetry \cite{ref35}. Since the
presence of the $\omega'_0 = 2$ resonance zone below the static
field ionization threshold does not allow direct static field
induced escape from the local chaotic band, the calculated
ionization probability $P_I = 21.3\%$ confirms the above view.

A quantum simulation at the same parameters gives a much lower
second peak at $n_s=73$ (see the dashed curve in Fig. \ref{fig4}
and notice the vertical logarithmic scale) and negligible
ionization probability, suggesting an (almost) adiabatic quantum
evolution. Even more suggestive is a comparison of  the evolution
of the classical and quantum population on the Stark levels during
the pulse: the two distributions are quite different even at early
times: quantum transfer of probability to levels around $n_s=73$
grows steadily during the rise of the pulse while in the classical
system we have a sudden spread of the probability after about $60$
microwave periods. At the peak of the pulse, classical and quantum
distributions on the Stark states again appear similar; but the
differences increase and become dramatic when, during the fall of
the pulse, we get to the second separatrix crossing and most of the
quantum population returns to the initial state. While it is only
from the second crossing on that we get a fundamental divergence
between classical and quantum behaviour, we do have significant
classical-quantum differences at both crossings, reflecting the
fact that it is near the separatrix that classical and quantum
systems are most different. At the separatrix a classical frequency
goes to zero \cite{ref36}, thus forbidding any adiabatic evolution
through the separatrix, but no quantum frequency (usually) goes to
zero, so that adiabatic evolution is possible \cite{ref37,ref53}.
Indeed at the peak of the pulse only two quasienergy states are
significantly populated: most of the probability ($94.6\%$)  is
still on the (separatrix) state ``adiabatically" connected to the
$n_0 = 65$ initial state; almost all of the rest ($5.1\%$) is on
the $n=77$ state (dashed line in Fig. \ref{fig2}) that right at the
peak of the pulse has a (narrow) avoided crossing with the $n_0 =
65$ state. As expected for an almost pure ``separatrix" state the
Husimi function of the wavefunction at the peak of the pulse is
localized over the separatrix region of the classical $\omega'_0  =
1/1$ resonance zone: see Fig. \ref{fig5} where the Husimi function
is superimposed over the classical surface of section. I tested
the stability of the state thus excited by inserting a long ($600$
microwave periods) flat central section to the microwave pulse.
Even if the state is not right at the separatrix (the Husimi
function of a true ``separatrix" state should have a peak centered
on the unstable fixed point, while in Fig. \ref{fig5} that peak has
already begun to split in two, each of these peaks closer to the
center of the resonance zone than the unstable fixed point), it
displays remarkable stability: a numerical fit of a two-terms
exponential decay model to the ionization probability during the
flat central region of the pulse gives a half lifetime of about
$2700$ microwave periods for the main component (about $95\%$)
against about $240$ microwave periods for the remaining $5\%$.

To check whether evolution on a single (long lived) quasienergy
state is the general behaviour for quantum states near $n_0 = 65$
at the same classical conditions, I have performed quantum
calculations at the same scaled parameters but at different values
of $n_0$ in the range $n_0 \in [57,73]$. In all cases we have
significant spread of the population at the peak of the pulse and
very sharp peaks at the end. Significant peaks away from $n_0$ (the
biggest one always on $n_s$, its ``symmetric" state with respect to
the center $n_r$ of the principal primary resonance \cite{ref38})
are visible only for $n_0 = 59$, $60$, $67$, $68$ and $73$. These
are also the only cases where significant ionization takes place as
can be seen in Fig. \ref{fig6}.a. These sharp ionization peaks as a
function of $n_0$ and the equally sharp peaks in the final
population distributions both tell us that for these values of
$n_0$ and of the peak microwave field $F^{max}$ (and we should also
include the pulse time $T$) quantum effects are essential: we are
still away from the hard semiclassical limit where multilevel
interactions dominate \cite{ref39}. Indeed Fig. \ref{fig6}.a can be
explained by two level quantum resonances: Fig. \ref{fig6}.b plots
(versus $n_0$ and in units of $\hbar \omega$) the difference
between the zero microwave field quasienergies of $n_0$ and $n_s$:
the ionization peaks correspond to the minima of this difference.
As a further test of the two level character of the interaction
responsible for the peaks, Fig. \ref{fig6}.c shows the very good
fitting of the Demkov model eq. (\ref{eqc1}) to the numerically
calculated population that has left $n_0$ at the end of the pulse.
For simplicity I have assumed the parameters $V_0$ and $B^2$ to be
$n_0$-independent so that eq. (\ref{eqc1}) reads $P^{(2)}_{a->b} =
K_1/ cosh^2( \Delta / K_2)$ where $\Delta$ is the quantity plotted
in Fig. \ref{fig6}.b and the two fitted parameters are $K_1 =
0.8337 \pm 0.0001$ and $K_2 = 0.01756 \pm 0.000005$. As a last test
of the above interpretation I have calculated the projection of the
wavefunction at the peak of the pulse on the instantaneous
quasienergy states for the case $n_0 = 67$ (that from Fig.
\ref{fig6}.b is one of the two minima of the difference between the
zero microwave field quasienergies of $n_0$ and $n_s$). The crosses
on the quasienergy curves in Fig. \ref{fig7} indicate the most
populated eigenstates: $39.1\%$ of the population is on the $n= n_0
=67$ state and $23.4+16.6\%$ on the state $n= n_s =75$ undergoing at that
field value a narrow avoided crossing with a ``second well" state
(dashed line). Using the above eigenstates as initial conditions
for integration of the Hamiltonian (\ref{eq1}) with $A(t)$ a
constant I obtained their half-lifetimes: the longest lived state
is the $n=67$ one (about $300$ microwave periods), the other two
states display an initial fast decay of their ``second well" part
(about $30\%$ of  the $n=75$ state and $70\%$ of the other decays
with a lifetime of about $10$ microwave periods) followed by a
decay corresponding to a lifetime of about $160$ microwave periods.
Both the peak microwave field quasienergy states $n=67$ and $n=75$
therefore have half lifetimes about one order of magnitude smaller
than the $n=65$ almost ``separatrix" state. This reduced lifetime
is likely to be reflected in Husimi functions of the eigenstates
considered. Moreover, as the main interaction is between two states
belonging to the principal primary quantum nonlinear resonance, it
is reasonable to assume that local characteristics of the Husimi
functions in the region of phase space of the principal primary
resonance zone will be relevant. Since the spacing of semiclassical
quasienergy curves quantized according to the resonance quantum
number $k$ is approximately uniform, whenever two zero microwave
field quasienergies are degenerate (or almost degenerate) the shift
of the two $k$-quantized quasienergy curves from the continuation
of n-quantized ones has maximum (see for example in fig 10 of ref.
\cite{ref17} the lengths of the segments connecting the inflection
points of the quasienergy curves calculated in the two
quantizations). This strong repulsion of the two quasienergies
corresponds to a substantial mixing of the two states; using the
terminology of ref. \cite{ref54} they are ``dirty".  The matter is
now to recognize a ``dirty" state from a ``clean" one. To this end
we note that ``quantum resonance" means that a quantum frequency is
zero; therefore, just like it happens at the separatrix in the
classical system, adiabatic evolution becomes impossible (or, in
the case of almost degeneracy, requires very long times). This
common characteristic of the two systems is reflected in the
somehow similar behaviour of the two systems we see for $n_0 = 59$,
$60$, $67$, $68$ and $73$. Since scarring of the separatrix state
is the one quantum states' characteristic associated (at least at
constant microwave field amplitude) with strong deviations from the
classical behaviour, it appears likely that the scarring of these
``dirty" states (when, at some time during the pulse they become
separatrix states) will be less pronounced than usual. Figure
\ref{fig8}.a shows for $n_0 =67$ the Husimi function of the
wavefunction at the peak of the pulse: it shows a very different
picture than the peak Husimi function  for $n_0 = 65$ (Fig.
\ref{fig5}). The maximum is still close to the classical unstable
fixed point, but the low, smooth ridge surrounding the classical
primary resonance zone is now marred  by the presence of two very
noticeable peaks. The Husimi functions of the three eigenstates
making up much of the instantaneous wavefunction (Fig.
\ref{fig8}.b) show that these two peaks are part  of the $n=n_0 =
67$ eigenstate. All three eigenstates have their maximum near the
unstable fixed point but as expected no eigenstate is exclusively
localized around the unstable fixed point. The closest we get to
this condition is with the Husimi function of the $n=n_s = 75$
eigenstate whose quasienergy, as we can see from Fig. \ref{fig7},
falls close to the inflection point of its quasienergy curve. The
quasienergy of the $n=n_0 = 67$ eigenstate instead appears to fall
to the right of the inflection point of its curve and accordingly
its Husimi function appears squeezed toward the center of the
principal primary resonance zone. Also, in this case, the ridge
surrounding the classical principal primary resonance zone is
absent from the Husimi functions of all three eigenstates. In its
stead the $n=n_s = 75$ eigenstate has two low peaks approximately
in the same position of the much higher peaks of the Husimi
function of the ``second well" eigenfunction. These peaks -in their
turn- alternate with the peaks of the $n=n_0 = 67$ eigenstate.
These correspondences and alternations (together with the knowledge
that away from avoided crossings the support of the ``second well"
state is  far from the principal primary resonance zone) lead us to
believe that in absence of the avoided crossing with the ``second
well" state, the ``separatrix" state would have (in the region of
the principal primary resonance zone) a Husimi function equal to
the sum of the $n=n_s = 75$ one and the ``second well" one. That
is: a wavefunction with noticeable peaks away from the unstable
fixed point. Our next step will be to try and locate, in the plane
$(\omega'_0,F_0^{max})$ of rescaled frequency $\omega'_0$ and peak
microwave field strength $F_0^{max})$, the range of parameters
where the adiabatic behaviour of the quantum states at the
separatrix we have just described results in the quantum ionization
probability being lower than the classical one. We call this
enhanced lifetime of the atom ``quantum stabilization".

\subsection{SIGNS OF QUANTUM STABILIZATION IN THE EXPERIMENTAL IONIZATION DATA}

\begin{figure}[htbp]
\centering\epsfig{file=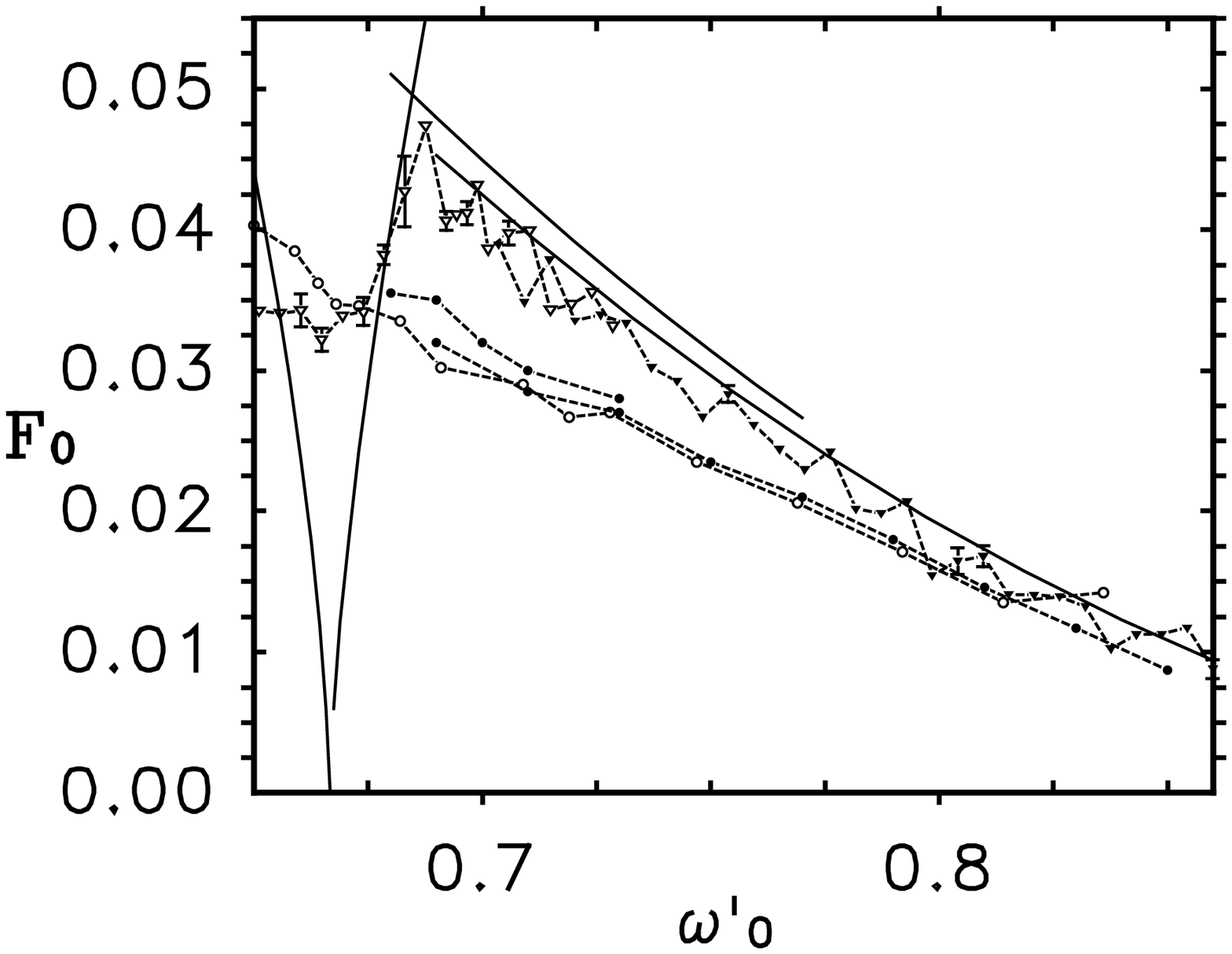,width=0.9\linewidth}
\caption{Signs of quantum localization in the ionization data.
Circles are the classical $F_0(10\%)$; triangles, full ($n_0 = 69$)
and empty ($n_0 = 65$), the experimental one. Heavy curves:
evaluation of the separatrix action location from eq. \ref{eqa3};
the ``V" on the left is for the $\omega'_0=2/3$ resonance zone, the
two curves on the right are for the principal primary resonance
zone, for the two cases $n_0= 65$, $F_S = 8\hspace{.1in} V/cm$
(upper curve)and $n_0 = 69$, $F_S = 8\hspace{.1in} V/cm$ (lower
curve). The experimental data for $\omega'_0>0.7$ are higher than
the classical ones and show reasonable agreement with the pendulum
approximation separatrix location curves. The full dots mark the
fields at which classical ensembles with initial conditions at
various values of $\omega'_0$ enter the primary resonance zone. (From Ref. \cite{ref40})}
\label{fig1}
\end{figure}

\begin{figure}[htbp]
\centering\epsfig{file=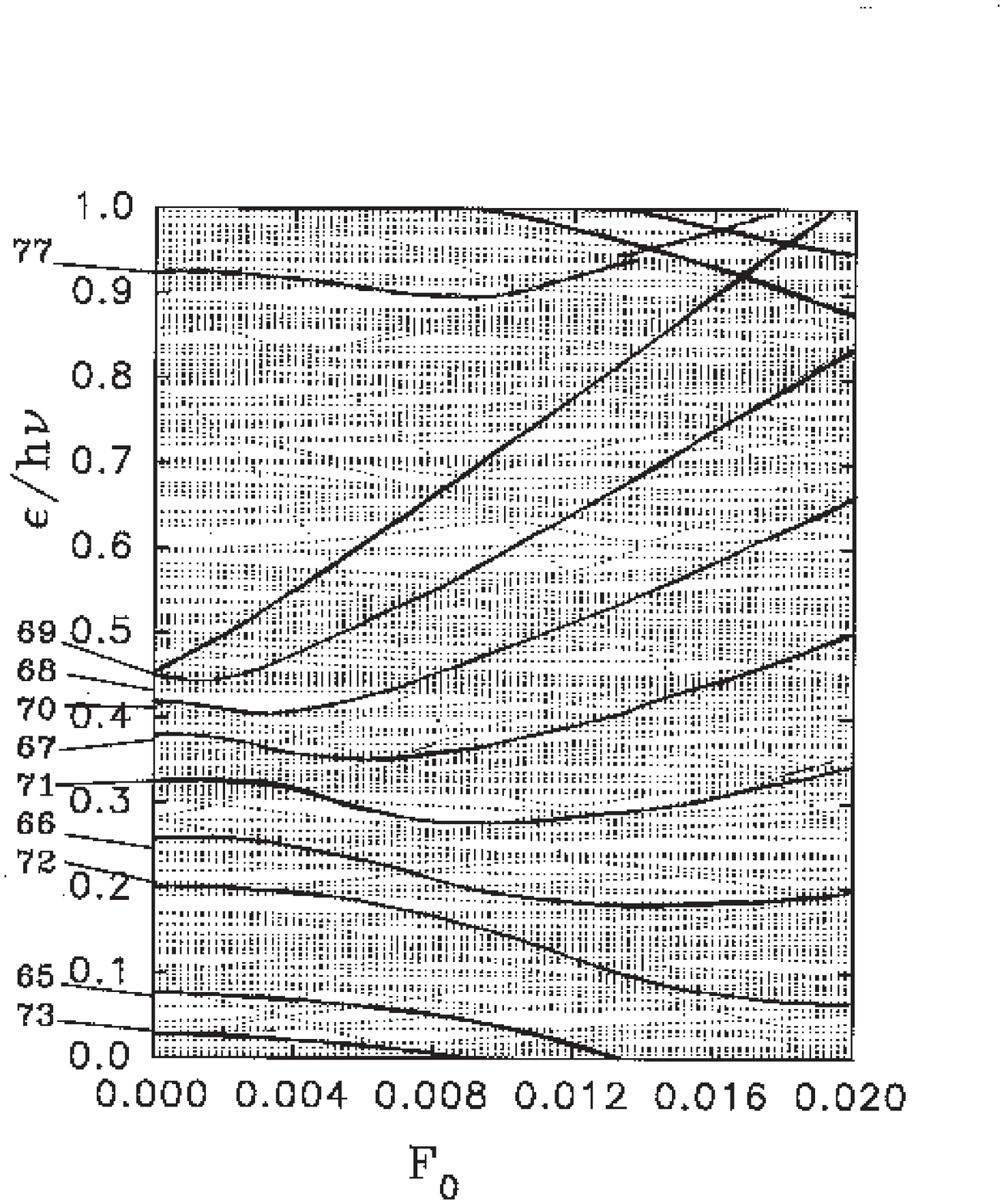,width=0.9\linewidth}
\caption{Quasienergy curves for the rescaled (to $n_0 = 65$) parameters $\omega'_0 = 0.8196$ and  $F_{0S} =0.02777$;
the horizontal scale is the rescaled microwave field strength. The
curves of the levels belonging to the $\omega'_0 = 1/1$ nonlinear
quantum resonance are marked as darker lines; the dashed line is
the $n=77$ level, having an avoided crossing with $n_0 = 65$ for
$F_0 \simeq 0.016$.}
\label{fig2}
\end{figure}

\begin{figure}[htbp]
\centering\epsfig{file=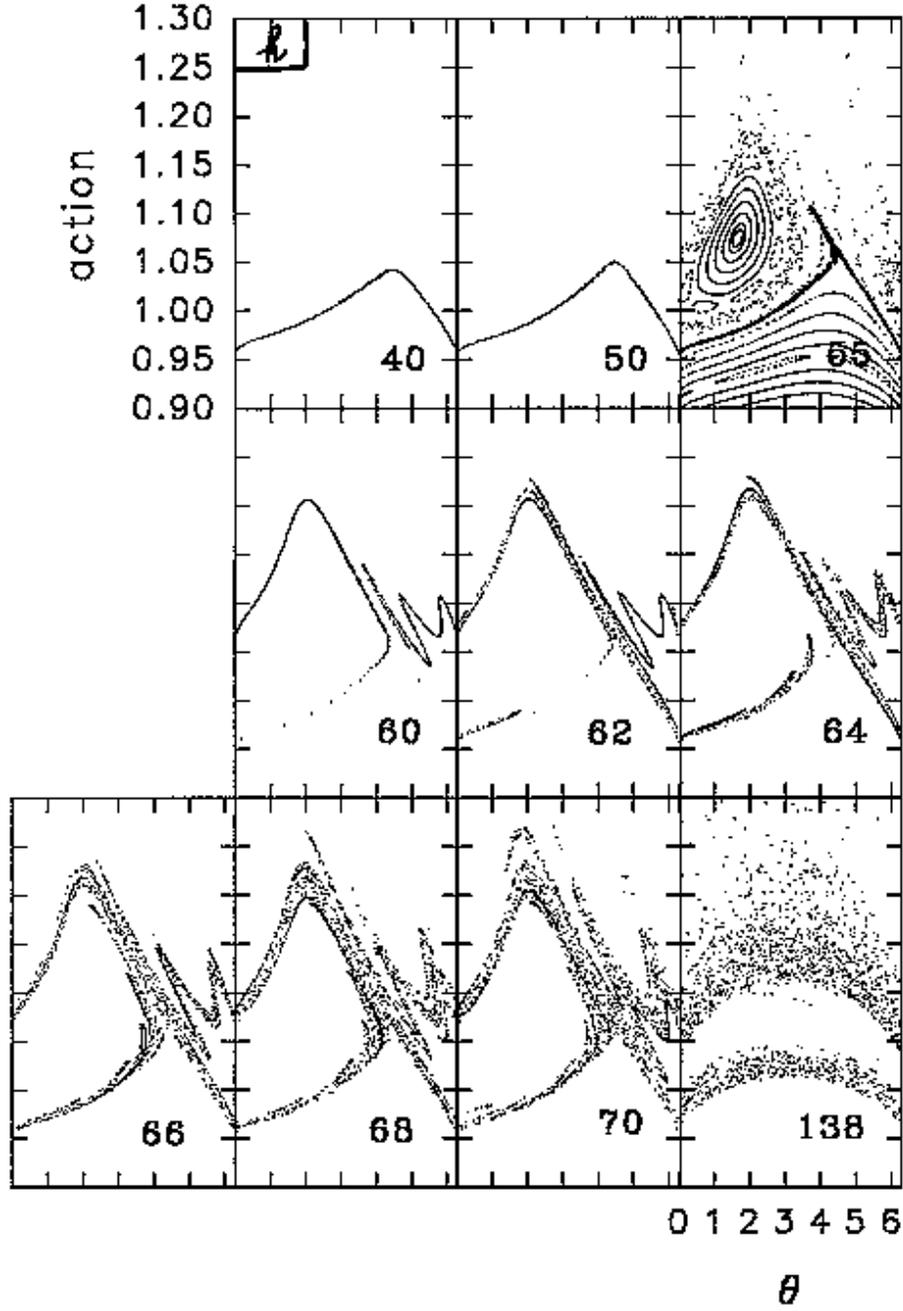,width=0.8\linewidth}
\caption{Classical ensemble at different times (indicated for each snapshot) during the rise of the
microwave pulse and at the end of it. The parameters are:
$\omega'_0 = 0.8196$, $F_0^{max} = 0.01607$, $F_{0S} =0.02777$, $T
= 140$ microwave periods. The size of Planck's constant $h$ for
$n_0 = 65$ is indicated. The instantaneous surface of section at
the crossing time is shown, displaying a wide chaotic separatrix
region. (From Ref. \cite{ref40})}
\label{fig3}
\end{figure}

\begin{figure}[htbp]
\centering\epsfig{file=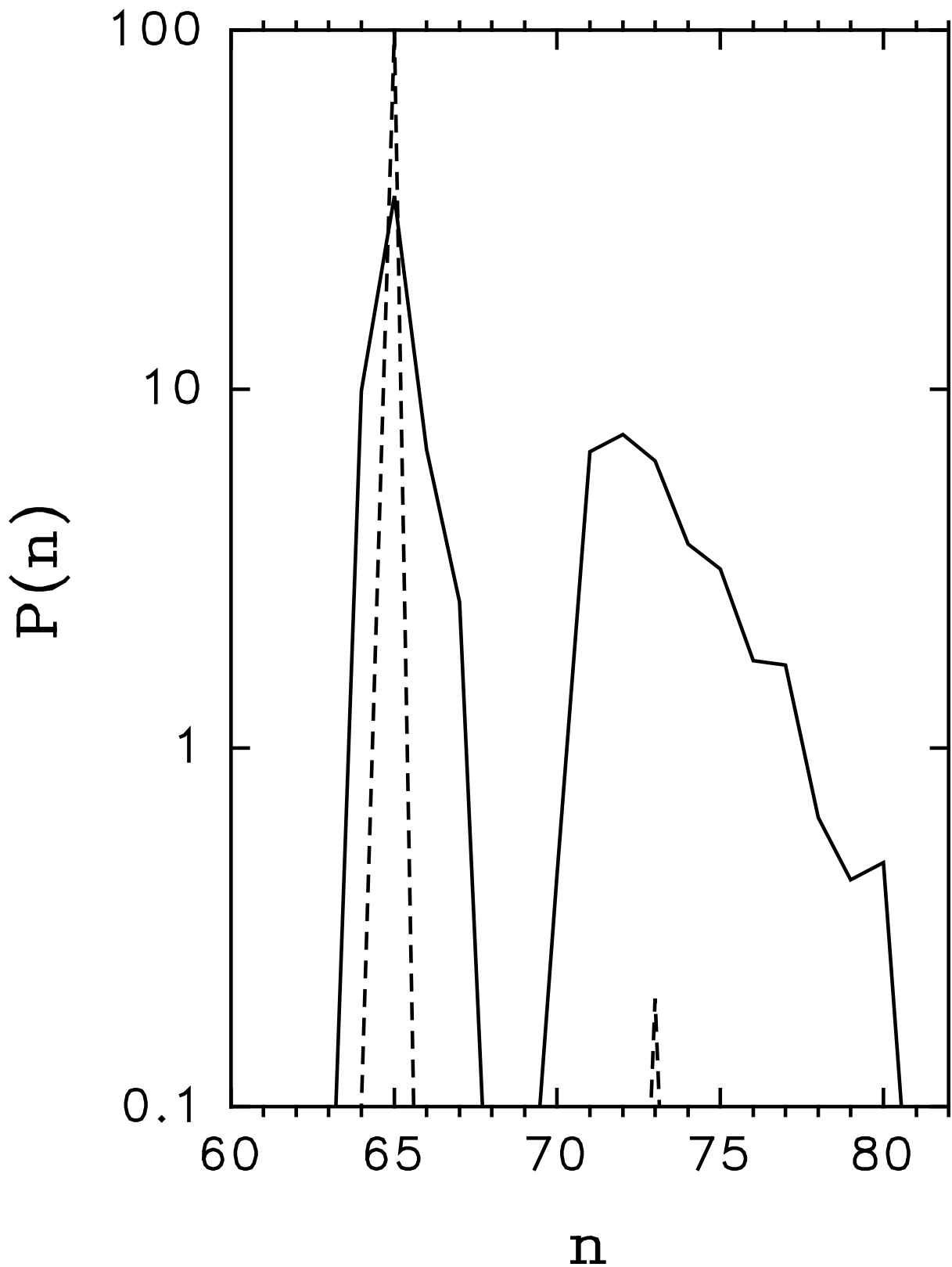,width=0.9\linewidth}
\caption{Classical (full line) and quantum (dashed line) final n-distribution for the
parameters $\omega'_0 = 0.8196$, $F_0^{max} = 0.01607$, $F_{0S} =0.02777$, $T
= 140$ microwave periods and $n_0 = 65$. The return of most of the
quantum population to $n_0 = 65$ suggests an (almost) adiabatic
evolution. (From Ref. \cite{ref40})}
\label{fig4}
\end{figure}

\begin{figure}[htbp]
\centering\epsfig{file=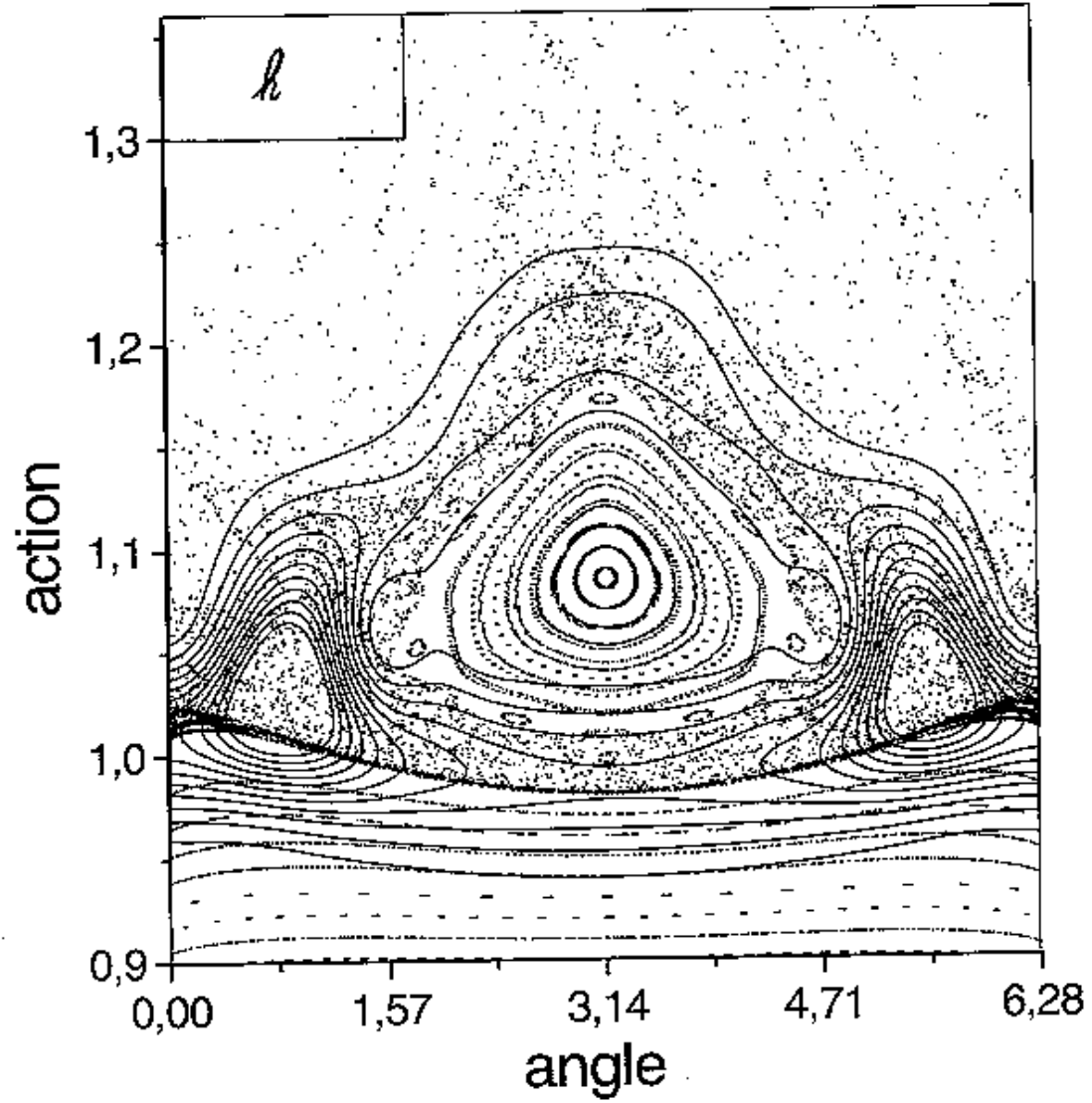,width=0.9\linewidth}
\caption{Husimi function of the quantum wavefunction at the peak of the pulse. The parameters are still
$\omega'_0 = 0.8196$, $F_0^{max} = 0.01607$, $F_{0S} =0.02777$, $T
= 140$ microwave periods and $n_0 = 65$. For comparison the classical instantaneous
surface of section is also shown. As suggested by the symmetric
shape of the Husimi function and its adherence to the outlines of
the classical surface of section, the wavefunction is almost
completely a single Floquet eigenstate.}
\label{fig5}
\end{figure}

\begin{figure}[htbp]
\centering\epsfig{file=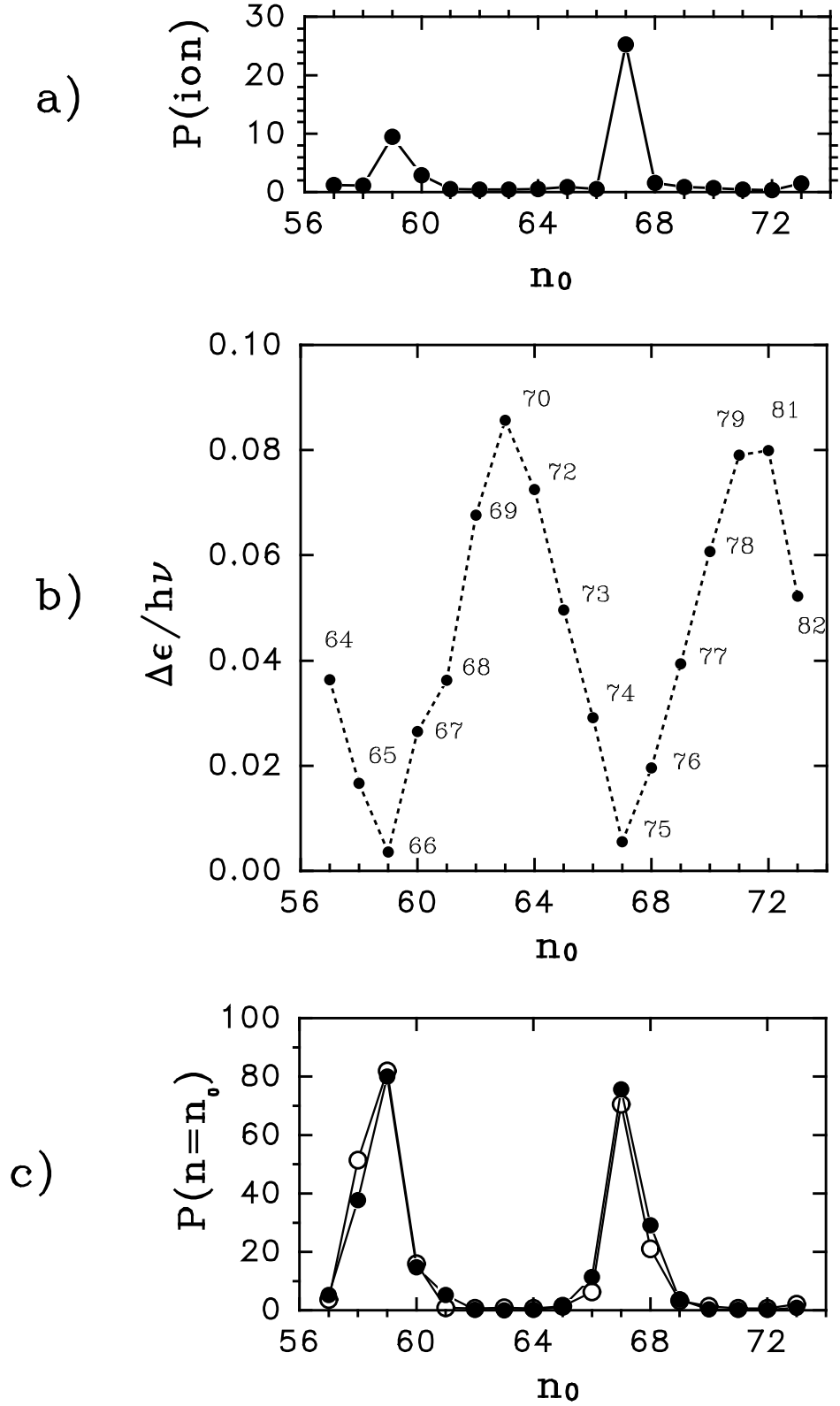,width=0.7\linewidth}
\caption{a) Quantum ionization probability versus initial Stark quantum number $n_0$. b) Zero microwave
field quasienergy difference between the initial quantum Stark
state $n_0$ and $n_s$ (indicated for every point) versus $n_0$. c)
fitting of the Demkov model (full circles) to the numerically
calculated population that has left $n_0$ at the end of the pulse
(open circles). The parameters are: $\omega'_0 = 0.8196$,
$F_0^{max} = 0.01607$ and $F_{0S} =0.02777$. Local maxima in a) and
c) correspond to minima in b) indicating the connection of
(semiclassical) nonadiabatic evolution with two level quantum
resonances. (From Ref. \cite{ref40})}
\label{fig6}
\end{figure}

\begin{figure}[htbp]
\centering\epsfig{file=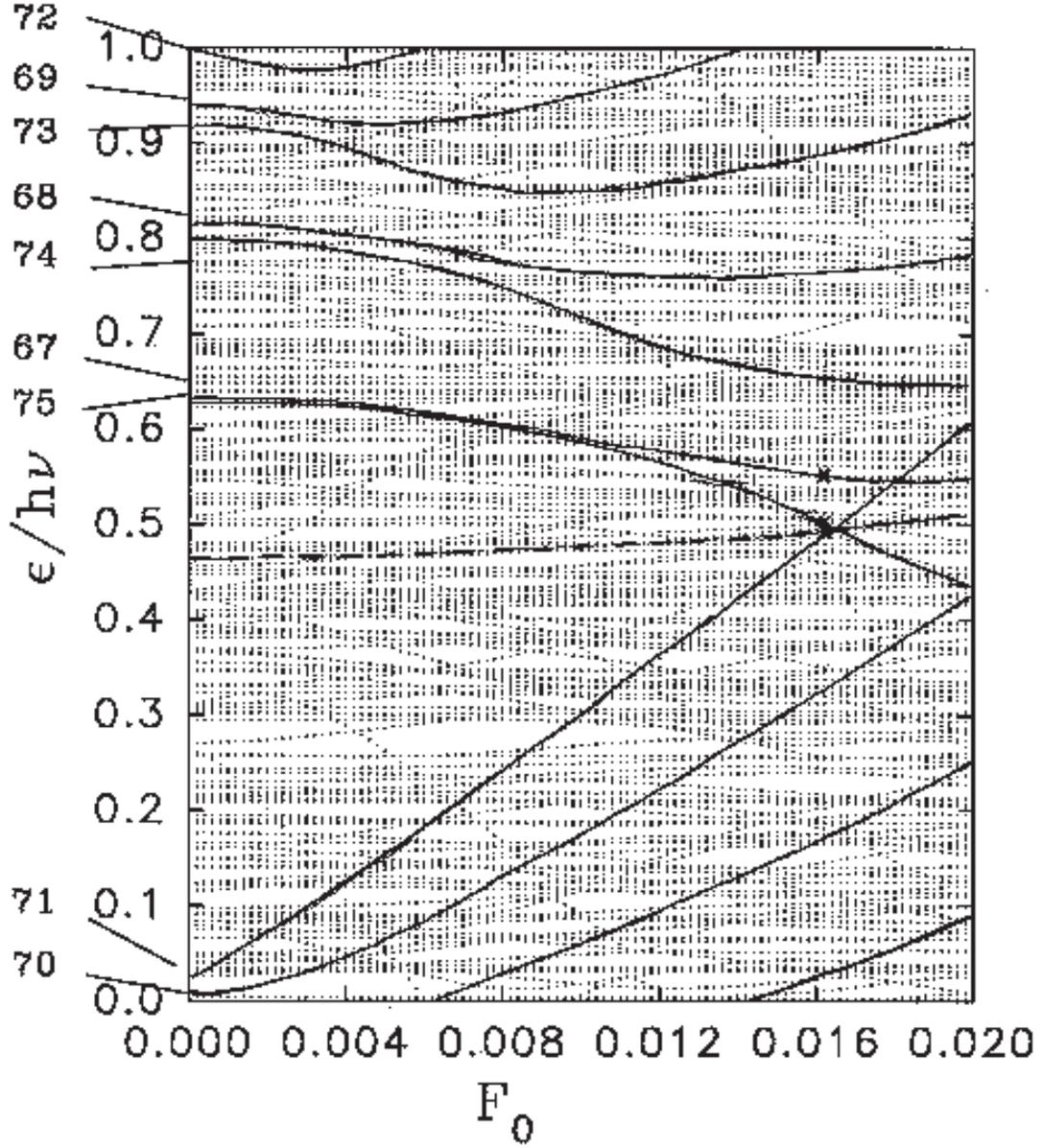,width=0.9\linewidth}
\caption{Quasienergy curves for the rescaled (to $n_0 = 67$) parameters $\omega'_0 = 0.8196$ and $F_{0S} =0.02777$;
the horizontal scale is the rescaled microwave field strength. The
curves of the levels belonging to the $\omega'_0  = 1/1$ nonlinear
quantum resonance are marked as darker lines. The crosses indicate
the quasienergies of the eigenstates most populated at the peak of
the pulse discussed in the text: $39.1\%$ of the population is on
the $n=67$ state and $23.4+16.6\%$ on the state $n=75$ undergoing
at that field value a narrow avoided crossing with a state of the
second well (dashed line).}
\label{fig7}
\end{figure}

Figure \ref{fig1} compares the scaled experimental peak microwave
field strength for $10\%$ ionization $F_0(10\%)$ (triangles)
\cite{ref17} with the same quantity from classical numerical
simulations (circles): starting from $\omega'_0\simeq  0.85$ and
down to $\omega'_0\simeq  0.60$ both curves show an increase in
$F_0(10\%)$ but the experimental $F_0(10\%)$ grows faster and the
sharp peak at $\omega'_0\simeq  0.69$ is clearly non classical. The
classical part of this increase can be easily explained: to be able
to ionize, the ensemble must first enter the chaotic ``separatrix
band" surrounding the principal primary resonance island. This can
be easily seen from a comparison of the classical $F_0(10\%)$ with
the fields at which classical ensembles with initial conditions at
various values of $\omega'_0$ enter the primary resonance zone
(dots in Fig. \ref{fig1}): the two are almost identical. Both of
the classical numerical curves in Fig. \ref{fig1} are lower than
the evaluation of the separatrix action location in asymmetric
pendulum approximation eq. \ref{eqa3} (heavy curves in Fig.
\ref{fig1}) that instead matches very well the $F_0(10\%)$
experimental data. The peak at $\omega'_0\simeq  0.69$ moreover
appears to be right at the intersection of the $\omega'_0=1/1$
separatrix location curve with the separatrix location curve of the
$\omega'_0
=2/3$ resonance zone. Eq. \ref{eqa3} is an approximation that, as
I explain in appendix A, only keeps the locally (in phase space)
resonant term in the Fourier expansion in the angle variable
$\theta$ of the interaction term between the electron and microwave
field. The agreement of the experimental data with it might
therefore be fortuitous; on the other hand I do suspect it to be
significant. Let us start by looking at some surfaces of section,
choosing the parameters along the experimental $F_0(10\%)$ curve.
Fig. \ref{fig9} shows the surface of section at a point to the far
right of Fig. \ref{fig1} where all the curves we have seen
approximately agree. The chaotic band is already rather well
developed and, as expected by the nonzero ionization probability,
already extends to high actions. On the other hand the manifolds of
the unstable fixed point still show oscillations only very close to
the fixed point itself (visible in the upper separatrix band on the
right hand side of the figure); it is therefore rather easy to draw
an approximate ``separatrix curve" that cuts through these
oscillations. It is immediately clear that while at the upper
separatrix the chaotic band extends on both sides of the
``separatrix curve", at the lower separatrix it is mostly confined
to the inner side. Therefore, if we approach the resonance zone
from below, entering the chaotic band will be equivalent to
crossing the ``separatrix curve". Fig. \ref{fig10}.a shows instead
a surface of section right at the $\omega'_0 \simeq 0.69$ peak: the
upper manifolds now oscillate wildly and drawing an approximate
``separatrix curve" becomes less intuitive \cite{ref16}, but it is
still possible to draw an intuitive approximate lower separatrix
curve (dashed line in Fig. \ref{fig10}.b). The inner side of the
chaotic band is still bigger, but now the outer side is
substantially thick (and merged with the chaotic band enveloping
the $\omega'_0 =2/3$ resonance zone). This time, approaching the
resonance zone from below,  we first enter the chaotic band and
then cross the ``separatrix curve". This explains in all likelihood
the disagreement of eq. \ref{eqa3} with the field for entrance into
the primary resonance zone, but what about the agreement of eq.
\ref{eqa3} with the experimental results? We know that the unstable
fixed point and its manifolds may support a scarred state and that
this state can be quite stable \cite{ref4}. The curve from eq.
\ref{eqa3} (obtained approximating these manifolds with a
``separatrix curve") gives us the position in the $(\omega'_0,
F_0)$ plane of this ``separatrix" state. Below that curve we expect
to find modulated free atom states; above, fast decaying states
with Husimi functions supported by the inner chaotic band
\cite{ref51}. The picture we have for our pulsed system is
therefore the following: for classical ionization to happen it is
enough to enter the chaotic band (assuming that at some point
during the pulse that same chaotic band merges with the local chaos
above), but for significant quantum ionization the quantum state
must evolve past the scarred separatrix state. The $\omega'_0
\simeq 0.69$ peak in the experimental $F_0(10\%)$ would then only
represent the ``point" in $\omega'_0$ where we pass from
selectively populating at the peak of the pulse the $\omega'_0=1/1$
separatrix state to populating some (less stable) state of the
$\omega'_0=2/3$ resonance zone; we shall see that this state is the
state at the center of the $\omega'_0=2/3$ resonance zone. The
quantum $F_0(10\%)$ for a number of cases is shown in Fig.
\ref{fig11}, compared with the corresponding classical and
experimental $F_0(10\%)$. Even if somehow lower in height and
shifted to lower frequencies ($\omega'_0 \simeq 0.682$ instead than
$\omega'_0 \simeq 0.687$) the quantum peak seems in reasonable
agreement with the experimental one. To confirm my supposition on
the origin of the $\omega'_0 \simeq 0.69$ peak, I have calculated
the projections of the wavefunction on the instantaneous Floquet
eigenstates at each period of the microwave during the first half
of the pulse for two cases along the quantum $F_0(10\%)$ curve at
the two sides of the peak: $\omega'_0 = 0.675$ and $0.685$; $F_0
= 0.04$ in both cases, but the field rescaled to the action at the
center of the resonance zone (see appendix D) is in the latter case
$2\%$ lower. They are shown in Figs. \ref{fig12}.a and
\ref{fig13}.a. Figs. \ref{fig12}.b and \ref{fig13}.b show instead
the quasienergy curves: I have marked as darker lines the
quasienergy curves of the $s/r=1/1$ nonlinear quantum resonance
(full lines) and (dashed lines) the two groupings of levels of the
$s/r=3/2$ one; the central state (in the free atom quantum number
$n$) of each grouping is the top one in the graph. For $\omega'_0 =
0.675$, immediately to the left of the quantum $F_0(10\%)$ peak in
Fig. \ref{fig1}, we see in Fig. \ref{fig12}.a that only at the peak
of the pulse the initially populated $n_0 = 65$ state (``central"
state of the top $s/r=3/2$ grouping and $k=16$ state of the
principal primary resonance) has a crossing with its nearest
$s/r=3/2$ state ($n=63$). Most of the population is therefore still
on a $s/r=3/2$ ``island" state as can be clearly seen from the
Husimi function at the peak of the pulse, shown in the top part
Fig. \ref{fig14}, superimposed over the classical surface of
section at that same time: the maxima of the Husimi function are
over the two classical $s/r=3/2$ islands. At the bottom  the same
figure are shown the Husimi functions of the three eigenstates on
which the instantaneous wavefunction has the highest projections:
most of the population ($62.3\%+15.7\%$) is on two states having
their maxima over the two islands of the $\omega'_0 =2/3$ resonance
zone (in the order: $n=65$ and the self ionizing state \cite{ref41}
having at that point a narrow avoided crossing with it) and only
$12.7\%$ is on the $n=63$ state, localized on the two unstable
fixed points of the same resonance zone. To the right of the peak
we instead see in Fig. \ref{fig13} that the $n_0=65$ and $n=63$
states cross at $F_0 \simeq 0.026$; at the peak of the pulse most
of the population is now on a state having the same character with
respect to the $s/r=1/1$ resonance  as in the case above but that
now is the second state of the top $s/r=3/2$ grouping. The maxima
of the Husimi function at the peak of the pulse are therefore no
more over the islands of the $\omega'_0 =2/3$ resonance zone but
have moved over to the two unstable fixed points as can be seen in
Fig. \ref{fig15}.a. The Husimi functions of the two eigenstates on
which the instantaneous wavefunction has the highest projections
(Figure \ref{fig15}.b) are again localized, one ($61.6\%$ of the
population) around the two unstable fixed points and the other
($19.3\%$ of the population) over the two islands of the $\omega'_0
=2/3$ resonance zone. In conclusion, at both sides of the $F_0(10\%)$
peak we have more than $60\%$ population on a single eigenstate but
their character is quite different as can also be seen from their
lifetimes. At $\omega'_0 = 0.685$ the most populated state is a
$\omega'_0 =2/3$ ``separatrix" state with a lifetime of about $210$
microwave periods while at $\omega'_0 = 0.875$ the most populated
state is a $\omega'_0 =2/3$ ``island" state with a lifetime of only
about $40$ microwave periods. It is interesting to note that while
the lifetime of the ``separatrix" state  only decreases from about
$210$ to about $180$ microwave periods when going from $\omega'_0 =
0.685$ to $\omega'_0 = 0.675$, the lifetime of the ``island" state
decreases from about $580$ microwave periods to only $40$. If we
attribute these decreased lifetimes to the slight increase the
microwave field strength rescaled to the action at the center of
the resonance we noted above, this behaviour would be in agreement
with the findings of ref. \cite{ref4} for the $\omega'_0 =1/1$
resonance zone: there it was noted that with increasing microwave
field strength the lifetime of the ``separatrix" states decreases
much more slowly than the lifetime of the ``island" states.

\section{CONCLUSIONS}

I have shown that, starting from a highly excited ($n_0 \simeq
65$) hydrogen atom and taking care to avoid some critical cases, it
is indeed possible to selectively populate one long-lived
``separatrix" state at the peak of a partially ionizing short pulse
of microwave electric field such as the ones used in the laboratory
experiments \cite{ref17} . This nonclassical quantum adiabatic
behaviour is due to the still low density of the resonance
quasienergy levels that allows the relevant quantum transition
frequencies to be much faster than the field strength change rate.
On the other hand, the spacing of these principal quantum nonlinear
resonance quasienergy levels for vanishing microwave field strength
is not uniform and care must be taken to avoid starting from a
state close to a (zero microwave field) two level quantum resonance
with one of its two nonlinear resonance neighbours. If the relevant
quantum frequency becomes comparable to the rate of change of the
field strength then adiabatic behaviour is no more possible: the
``local" increase in the density of resonance states has made the
system more semiclassical. I have also shown that the sharp
$\omega'_0 \simeq 0.69$ maximum in $F_0(10\%)$ (on which all
previous studies dealing with this form of quantum stabilization
\cite{ref11,ref18,ref19,ref20} concentrated) only marks the
``crossover" point from the $\omega'_0=1/1$ resonance zone to the
$\omega'_0=2/3$ one. It is instead the whole region $\omega'_0 \in
(0.69, 0.85)$ that shows clear signs of the quantum stabilization
induced by the selective population of the separatrix state: with
decreasing frequency the divergence between the quantum (and
experimental) and the classical $F_0(10\%)$ increases. I have
shown that this divergence is due to the increase in the width of
the classical chaotic ``separatrix" band: for classical ionization
to happen it is enough to enter the chaotic band, but for
significant quantum ionization the quantum state must evolve past
the scarred separatrix state. The stability of the separatrix state
itself appears instead from my calculated lifetimes to decrease
with decreasing frequency.

\section{Aknowledgements}

The author wishes to thank C. Rovelli, G. Mantica and S. Locklin for
useful comments and suggestions and the latter also for the use of
his computer for the quantum numerical simulations presented in
this paper. Special thanks to J.E. Bayfield as advisor of my Ph. D. thesis, amply quoted in the present paper.

\appendix
\section{classical pendulum approximation for the separatrix location curve \cite{ref40}}
\label{app:a}

When $A(t)$ is a constant, one procedure often used to study the
Hamiltonian (\ref{eq1}) is to locally approximate it with an
integrable one around any resonant action Ir defined by $\omega
I_r^3/(1-3F_S I_r^4) = s$ where $s$ is an integer \cite{ref25}. In
brief: we first substitute the atom-static field interaction term
with its average over one free atom period \cite{ref42}. Then, the
free atom Hamiltonian is expanded in powers of $x=(I - I_r)$ up to
the second order term and the atom-field interaction term is
Fourier expanded in the angle variable $\theta$. All terms of the
latter expansion but the $s$-th one average to zero over one period
of the microwave field; they can therefore be neglected in first
approximation. The fast motion with frequency $\omega$ can be
extracted by a canonical transformation canceling the time
dependence of the interaction term; this will also cancel the
linear term in the expansion of the free atom Hamiltonian thus
leaving us with the Hamiltonian of a pendulum describing the slow
motion inside the resonance island
\begin{equation}
H^{(2)}  =  - {\frac {3(1+F_SI_r^4)} {2I_r^4}}x^2 - F I_r^2 {\frac {J'_s(s)} s}\cos{s\theta}
\label{eqa1}
\end{equation}
where $J's(s)$ is the derivative of the Bessel function $J$. The
above Hamiltonian highlights the component of the interaction
responsible for the most noticeable structures in phase space,
namely the primary island chains corresponding to free atom
rotation numbers $1/s$. As costumary, we shall use the
approximation $J'_s(s)/s \simeq \mu /s^{5/3}$ \cite{ref43}, where
$\mu
= 0.325$ if $s=1$ and tends to $\mu = 0.411$ for $s \to
\infty$. From the Hamiltonian \ref{eqa1} it is then easy to obtain
the equation in phase space of the separatrix between rotational
and librational motion. Since the action associated with an
invariant curve $I(\theta)$ is the integral in $\theta \in [0,
2\pi)$ of $I(\theta)$ divided by $2\pi$, calling ${\mathcal
A}^{\pm}$ the two integrals for the upper and lower separatrix, the
two actions
\begin{eqnarray}
I^{\pm}= I_r \pm  {\frac {{\mathcal A}^{\pm}} {2\pi}}\nonumber
\end{eqnarray}
are the actions just above the upper separatrix ($I^+$) or just
below the lower separatrix ($I^-$).

In the limit of very slow rise of the pulse the action is an
invariant; in that limit $I^{\pm}$ therefore represent, for an atom
in a pulsed microwave field with peak microwave field strength $F$,
the free atom actions of points that are at the separatrix at the
peak of the pulse. We call them the ``separatrix actions" of the
resonance zone. This derivation is discussed in ref. \cite{ref19};
but there the interaction term of the pendulum Hamiltonian contains
a higher order correction to the atom-microwave interaction term
that I think inconsistent with a first order approximation in both
$F$ and $F_S$. A modified pendulum approximation for the slow
component of the regular motion inside the resonance island can be
used to take into account the asymmetry in action of the resonance
zone; in this new approximation the expansion of the $-1/2I^2$ term
in the Hamiltonian around the resonant action $I_r$ is carried one
term further \cite{ref44}. Including both the first order static
field correction and the asymmetry term, my calculation yields the
following formula in the case of the $\omega'_0 = 1$ resonance
zone:
\begin{eqnarray}
{{I^{\pm}}{I_r}}= 1 \pm {\frac 4 \pi}{(BF_r)}^{1/2} + {\frac 4 3}{\frac {BF_r} {(1+F_S^r)}},\nonumber
\end{eqnarray}
where $B = 0.325/[3(1+F_S^r)]$. Here $F_S^r$ is the static field
scaled by $I_r^4$. This equation can be used to obtain a
``separatrix location function" in the $(\omega'_0,F_0)$ parameter
space which, expressed in terms of the parameter $u_{\pm}= I_r
/I^{\pm}$, is:
\begin{eqnarray}
\omega'_0{}^{\pm} = s {\frac {1 - 3 F_{S0} u_{\pm}^4} {(1 - 3 F_{S0})u_{\pm}^3}}\hspace{1.0in}\nonumber\\
F_0^{\pm} = \left({\frac {1- \left[1 - {\mathcal W}\left(1
-{\frac 1 {u_{\pm}}}\right)\right]^{1/2}} {\mathcal WV}}\right)^2
\label{eqa3}
\end{eqnarray}
where
\begin{eqnarray}
{\mathcal V}= {\frac 2 {\pi}} \left[{{\frac {\mu u_{\pm}^4} {3s^{5/3}(1 + F_{S0}
u_{\pm}^4)}}}\right]^{1/2}\nonumber
\end{eqnarray}
and
\begin{eqnarray}
{\mathcal W}= {\frac{\pi^2} {3(1 + F_{S0} u_{\pm}^4)}}.\nonumber
\end{eqnarray}

\section{Classical and quantum change of behaviour at the separatrix \cite{ref40}}
\label{app:b}

To stress the importance of the inflection point of the quasienergy
curves as the transition point between two very different types of
quasienergy states, ref. \cite{ref45} calculated the
time-independent expectation value of the unperturbed Hamiltonian
$H_0$ on the quasienergy states themselves:
\begin{equation}
H_0^{k,k} = < \phi_k(z,t)  | H_0 | \phi_k(z,t) > = \Sigma_n
a_{kn}^2 E_n
\label{eqb1}
\end{equation}
there is a fast change of behaviour at a value of $k$ that in each
case is in good agreement with $k_{max} = 8(0.325 F/3)^{1/2}
/\pi\omega - 1/2$. To the right of that point $H_0^{k,k} \simeq E_{n(k)}$ (the unperturbed
energy of the corresponding $n(k)$), meaning that the only
coefficients in eq. (\ref{eqb1}) significantly different from zero
are $a_{kn(k)}$ and the $a_{kn}$ with $n$ close to $n(k)$, that is:
only states on the same side of the resonant state $R$ give a
significant contribution in (\ref{eqb1}). To the right, $H_0^{k,k}$
decreases with increasing $k$ but remains close to the energy $E_R$
of the resonant state: the significant coefficients $a_{kn}$ in the
sum (\ref{eqb1}) correspond now to $n$'s both below and above $R$.
This behaviour is present also in the classical description and the
two limiting behaviours can be easily understood as follows:
outside of the resonance zone the motion is still very similar to
the unperturbed one, so that the average atomic energy $<H_0>$ for
a given orbit remains close to the free atom one. Inside the
resonance zone the motion is the sum of a fast component, that is
the motion of the stable resonant orbit at the center of the
resonance zone, and of a slow one around that same orbit resulting
in the weak dependence of the energy $H_0^{k,k}$ on $k$. The slow
motion determines in the quantum picture the spatial structure of
the states, the fast one appears instead as oscillations in time,
locked to the driving field frequency and in phase with the
oscillations of the stable periodic orbit at the center of the
resonance zone. It is possible to formalize this intuitive
classical picture as follows: we again make use of the pendulum
Hamiltonian (\ref{eqa1}) and to keep the notation light we pass to
rescaled variables and limit ourselves to the main resonance $s=1$:
\begin{eqnarray}
H  =  - \beta {\frac{x^2} 2} +
\alpha\cos{(\theta)}\hspace{1.4in}\nonumber\\
\beta=3(1+F_SI_r^4)\hspace{.2in}\alpha = 0.325 F_0\nonumber
\end{eqnarray}
With this notation the average on a given orbit of the free atom energy is:
\begin{eqnarray}
\left<H_0\right> = -\left<{\frac 1 {2 I^2}}\right> \simeq  -{\frac 1 {2 I^2}}+ {\frac{\left<x\right>} {I_r^3}} + \left<H_0^{(1)}\right>\nonumber
\end{eqnarray}
where $H_0^{(1)} = - \beta x^2/2$ and the averages are calculated
over the pendulum invariant curves. Both $\left<x\right>$ and
$H_0^{(1)}$ change markedly when passing from inside to outside the
separatrix. Inside the separatrix we have
\begin{equation}
\left< x \right> = 0
\label{eqb2a}
\end{equation}
and outside (the + sign will of course be for orbits above the separatrix and vice versa):
\begin{equation}
\left<x\right> = \pm \pi(\alpha/\beta)^{1/2}/R{\bf K}(R)
\label{eqb2b}
\end{equation}
where  $R = ( 2/(1 - H/ \alpha))^{1/2} = 1/|sin{(\Theta/2)}|$,
$\Theta$ being the (half) amplitude in $\theta$ of the
oscillations, and ${\bf K}$ is a complete elliptic integral
\cite{ref46}. Expression \ref{eqb2b} is zero for $R=1$(at the
separatrix) and for $R \to 0$ becomes
\begin{equation}
\left<x\right> = \pm (2|H_0|/\beta)^{1/2}\equiv \pm J
\label{eqb2c}
\end{equation}
The transition happens very fast and close to the separatrix:
already at $R=0.95$ the ratio $|\left<x \right>/J|$ is about
$0.87$. Outside the separatrix, $\left<H_0^{(1)}\right> =
-J\pi(\alpha\beta)^{1/2}/2R{\bf K}(R)$ similarly shows a sudden
transition: it is again zero at the separatrix and for $R\to 0$ it
becomes
\begin{equation}
\left<H_0^{(1)}\right> \simeq - |H_0| = -\beta J^2/2;
\label{eqb2d}
\end{equation}
but already at $R=0.95$ the ratio $|\left<H_0^{(1)} \right>/ (\beta
J^2/2)|$ is about $0.87$. Inside the separatrix
$\left<H_0^{(1)}\right> =
-J\pi(\alpha\beta)^{1/2}/4{\bf K}(1/R)$ that for $1/R \to 0$ goes as
\begin{equation}
\left<H_0^{(1)}\right> \simeq -J(\alpha\beta)^{1/2}/2,
\label{eqb2e}
\end{equation}
has a minimum for $1/R \simeq 0.94$, where the ratio
$\left<H_0^{(1)}\right>/ (\beta(J/2)^2/2)$ is about $-1.205$ and
becomes zero at the separatrix. Summarizing, $\left<x\right>$ grows
suddenly when crossing from inside to outside the separatrix,
$\left<H_0^{(1)}\right>$ has a sharp dip at the separatrix itself.
Far from the resonance zone we shall therefore have from eqs.
(\ref{eqb2d}) and (\ref{eqb2e})
\begin{eqnarray}
\left<H_0\right> \simeq -{\frac 1 {2 I_r^2}}\pm {\frac J {I_r^3}} - |\Delta_k(0)|
= -{\frac 1 {2 I_r^2}}\pm J\omega - |\Delta_k(0)|;\nonumber
\end{eqnarray}
it is now sufficient to remember that $|\Delta_k(0)| = | E_n - E_R
- (n-R)\omega|$ and notice that, outside of the resonance zone, $J = |n
-R|$, to obtain, in agreement with the quantum result in ref.
\cite{ref45}, $\left<H_0\right> \simeq E_n$. Close to the center of
the resonance zone we shall instead have, from eqs. (\ref{eqb2a})
and (\ref{eqb2c}):
\begin{eqnarray}
\left<H_0\right> \simeq -{\frac 1 {2 I_r^2}}- {\frac {k +1/2} {2I_r}}( 0.975F)^{1/2}\nonumber
\end{eqnarray}
that for each of the three figures in ref. \cite{ref45} gives us
slopes in good agreement with the numerical quantum results. The
main difference between inside and outside the resonance zone, and
the cause of the sudden jump in the behaviour of $H_0^{k,k} $, is
clearly the term in $\left<x\right>$ that is exactly zero inside
the resonance zone (at least in pendulum approximation) and has a
finite value immediately out of it. Its being zero inside the
resonance zone is an expression of that locking of the resonance
states' oscillation frequency to the microwave one I mentioned
above \cite{ref47}.

\section{The Demkov model}
\label{app:c}

The Demkov model \cite{ref48} deals with the interaction of two
levels and assumes the two diabatic levels to be constant and the
interaction between them to change exponentially with the
perturbation parameter $\lambda$. Assuming the time to vary between
$-\infty$ and $+\infty$ the Hamiltonian matrix will have the form
\begin{eqnarray}
H =\left[{\begin{array}{cc}
    H_{11} & H_{12} \\
    H_{21} &H_{22}
  \end{array}}\right]=\left[{
  \begin{array}{cc}
    H_{11} (|t|) & V_0e^{-B^2|t|} \\
    V_0e^{-B^2|t|} & H_{11} (|t|) - |\Delta|
  \end{array}}\right]\nonumber
\end{eqnarray}
and the levels will behave as in Fig \ref{figc1}. For $t = \pm
\infty$ the adiabatic levels will coincide with the diabatic ones ($a$
with $1$ and $b$ with $2$); the model moreover assumes $|V_0| \gg |
\Delta |$ so that at $t=0$ the adiabatic levels will be approximately $H_{11}(|t|)
- | \Delta |/2 \pm |V_0| $ (the $+$ sign for level $a$) and the corresponding states symmetric
(state $a$) and antisymmetric (state $b$) combinations of the two
diabatic ones. For a complete pulse ($t$ going from $-\infty$ to
$+\infty$) the transition probability between the two states is
\begin{equation}
P^{(2)}_{a->b} =  [sin (2V_0/\hbar B^2)/ cosh(\pi
\Delta/2\hbar B^2)]^2
\label{eqc1}
\end{equation}
The model assumes an infinite pulse time, on the other hand the
result above can be thought to give an approximate transition
probability for actual finite pulses; for this it is convenient to
rewrite it in a more general form. $V_0$ represents the maximum
value $H_{12}^{max}$  reached by the interaction term $H_{12}$ at
the peak of the pulse; $B^2$ can instead be written as
$|dH_{12}/dt| / H_{12}$. Since nonadiabatic transitions are
expected to take place when $H_{12}\approx |\Delta|$ (that is when
the separation of the adiabatic levels is about twice the
separation $|\Delta|$ of the diabatic ones) \cite{ref48},  to apply
the model to a case where $|dH_{12}/dt|$ is not a constant, our
best choice will be to calculate $B^2$ when $H_{12} = |\Delta|$. A
classical resonance zone in phase space that is not only always
present in a certain region but also keeps growing with increasing
values of the perturbation parameter, can manifest itself in a
Demkov-like interaction of quantum levels \cite{ref49}. The Demkov
model can therefore be useful when dealing with two level
interactions related to classical primary resonances.

\section{scaling \cite{ref40}}
\label{app:d}

Figure \ref{figd1}.a shows, the experimental peak microwave field
strength for $10\%$ ionization $F_0(10\%)$ vs. $\omega'_0$, in the
region $\omega'_0 \approx 1$ extended beyond the central peak of
classical stability due to trapping of the ensemble within the
principal primary resonance island. We observe a gradual increase
of $F_0(10\%)$, very clear for decreasing frequencies and less so
on the high frequency side. This lack of clarity is partly due to
the big step we encounter between the $n_0=72$ and the $n_0=80$
data: to keep the ionization cutoff in $n$ (usually around $n=90$
for $n_0=65,69,72$) far enough from $n_0$, the static field was
changed from $8\hspace{.1in}V/cm$ to $1\hspace{.1in}V/cm$ in the
$n_0=80$ experiments (so that the ionization cutoff is moved to $n
\simeq 150$). Adapting to the present case the suggestion of mixed units
advanced in ref. \cite{ref50} Figure \ref{figd1}.b replots the data
as follows: the frequency is still rescaled to $n_0$, but the field
strength is rescaled to the action at the center of the resonance
and corrected to the first order for the static field $F'_r
= F_r(1 +7 F_{rS})$ \cite{ref17}. The data now connect reasonably well
but since the new vertical scale is stretched with respect to $F_0$
for $\omega'_0 < 1$ and compressed for $\omega'_0 > 1$ (and
increasingly so the further we move away from $\omega'_0 =1$), it
progressively flattens the ionization threshold structures the
higher the values $\omega'_0$ at which they appear. Finally we note
that if $\omega'_0$ is changed by changing $n_0$ (as it is the case
in ref. \cite{ref50}), then the ratio between $F$ and $F'_r$ is
constant; if $\omega'_0$ is instead changed by changing $\omega$
itself (as it is the case for each of the four series of data in
Fig. \ref{figd1}), it is the ratio between $F$ and $F_0$ that
remains constant.



\newpage
.

\newpage
.

\newpage
.

\begin{figure}[htbp]
\centering\epsfig{file=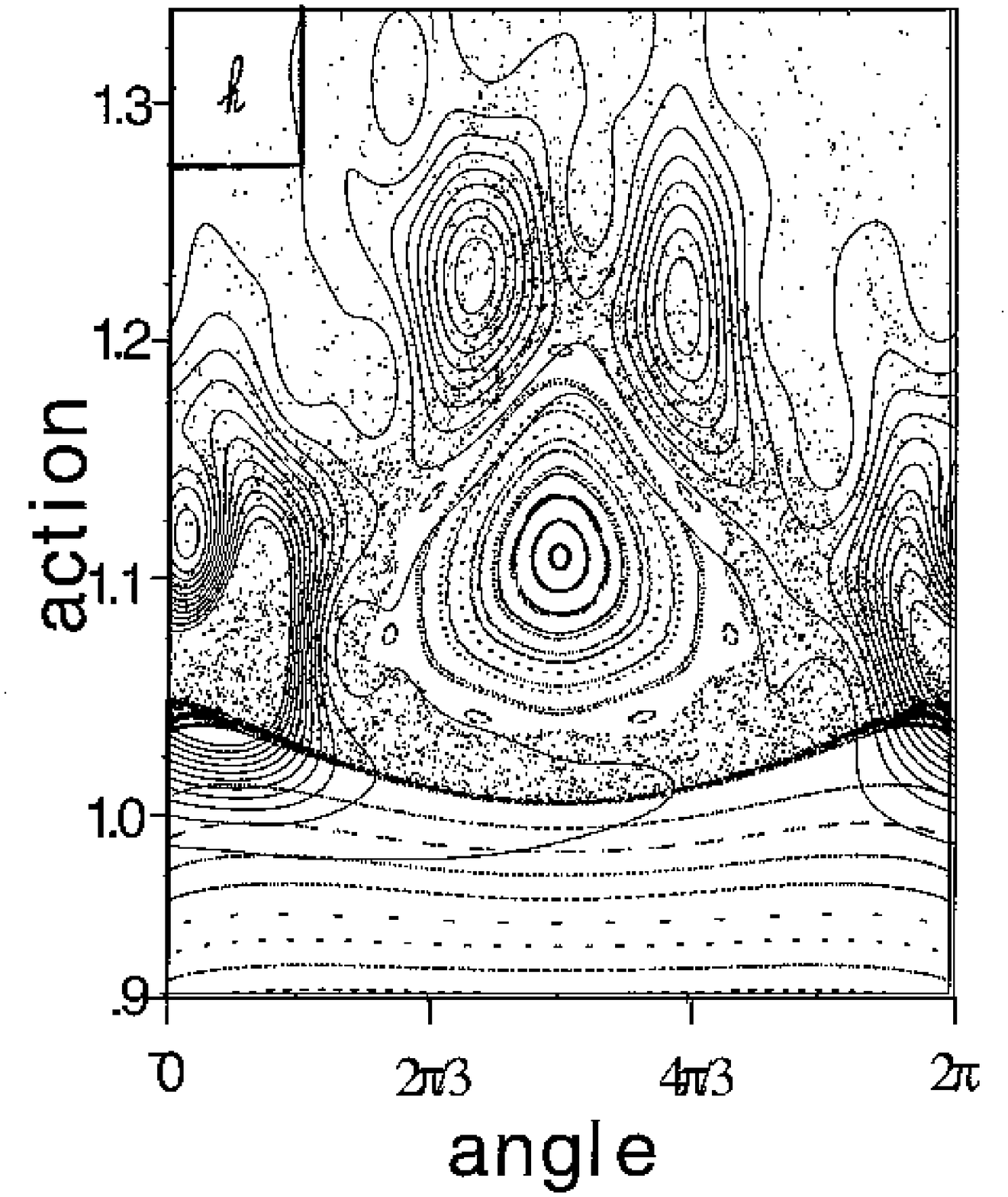,width=0.9\linewidth}
\end{figure}

\begin{figure}[htbp]
\centering\epsfig{file=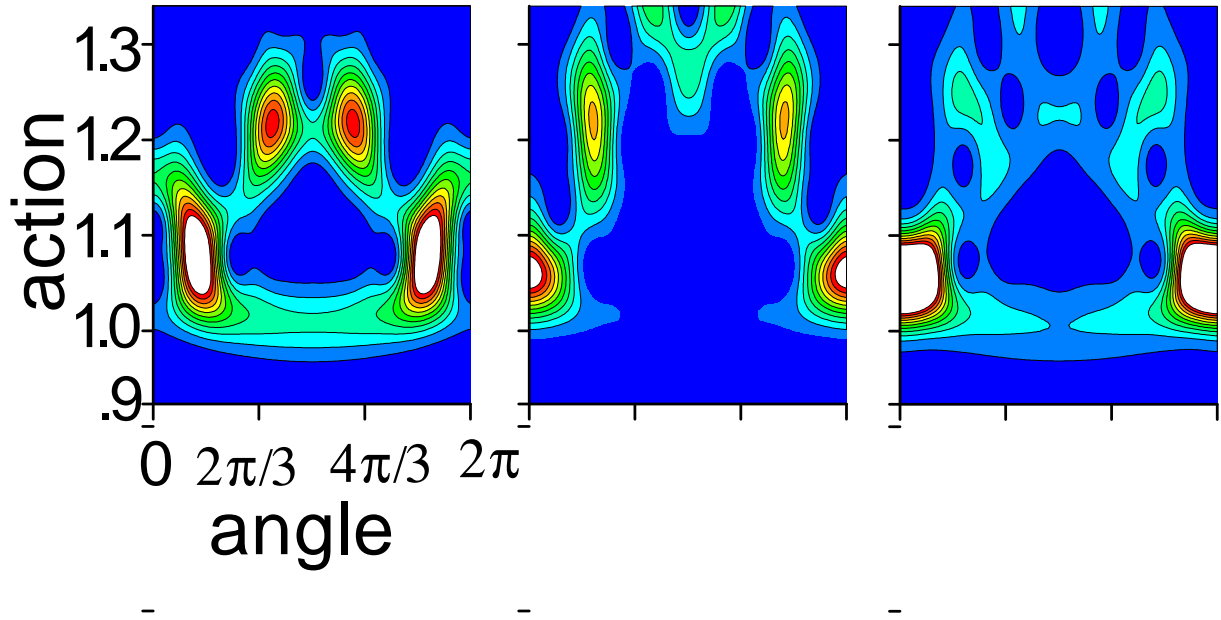,width=0.9\linewidth}
\caption{a) Husimi function of the quantum wavefunction at the peak of the pulse. The parameters are
$\omega'_0 = 0.8196$, $F_0^{max} = 0.01607$, $F_{0S} =0.02777$, $T
= 140$ microwave periods and $n_0 = 67$. For comparison the
classical instantaneous surface of section is also shown. The
symmetry of Fig. \ref{fig5} has been broken and the Husimi function
can only losely be said to reflect the classical phase structure in
its avoidance of the principal primary island at the center of the
figure. b) Husimi functions of the three eigenstates on which the
above instantaneous wavefunction has the highest projections. In
the order: the $n_0 = 67$ state ($39.1\%$), the $n_s
= 75$ state ($23.4\%$) and the ``second well" state undergoing a narrow avoided crossing with the latter
($16.6\%$).}
\label{fig8}
\end{figure}
\newpage
.

\begin{figure}[htbp]
\centering\epsfig{file=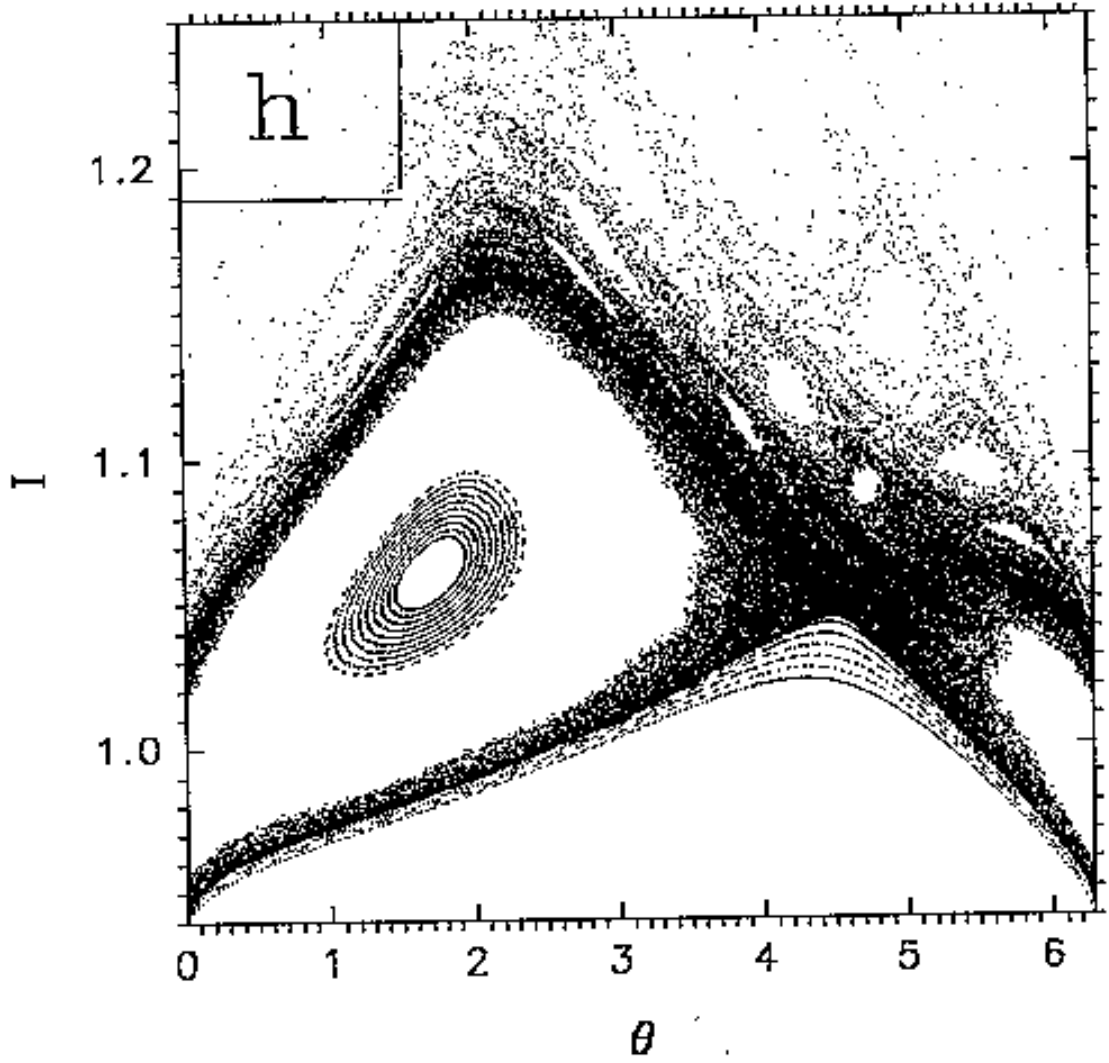,width=1.0\linewidth}
\caption{Surface of section (Poincar\'{e} map) for $\omega'_0 = 0.847$, $F_0 =
0.0104$, $F_{0S} = 0.035266$ and $\phi_0 = 0$. (From Ref. \cite{ref40})}
\label{fig9}
\end{figure}
\newpage
.

\begin{figure}[htbp]
\centering\epsfig{file=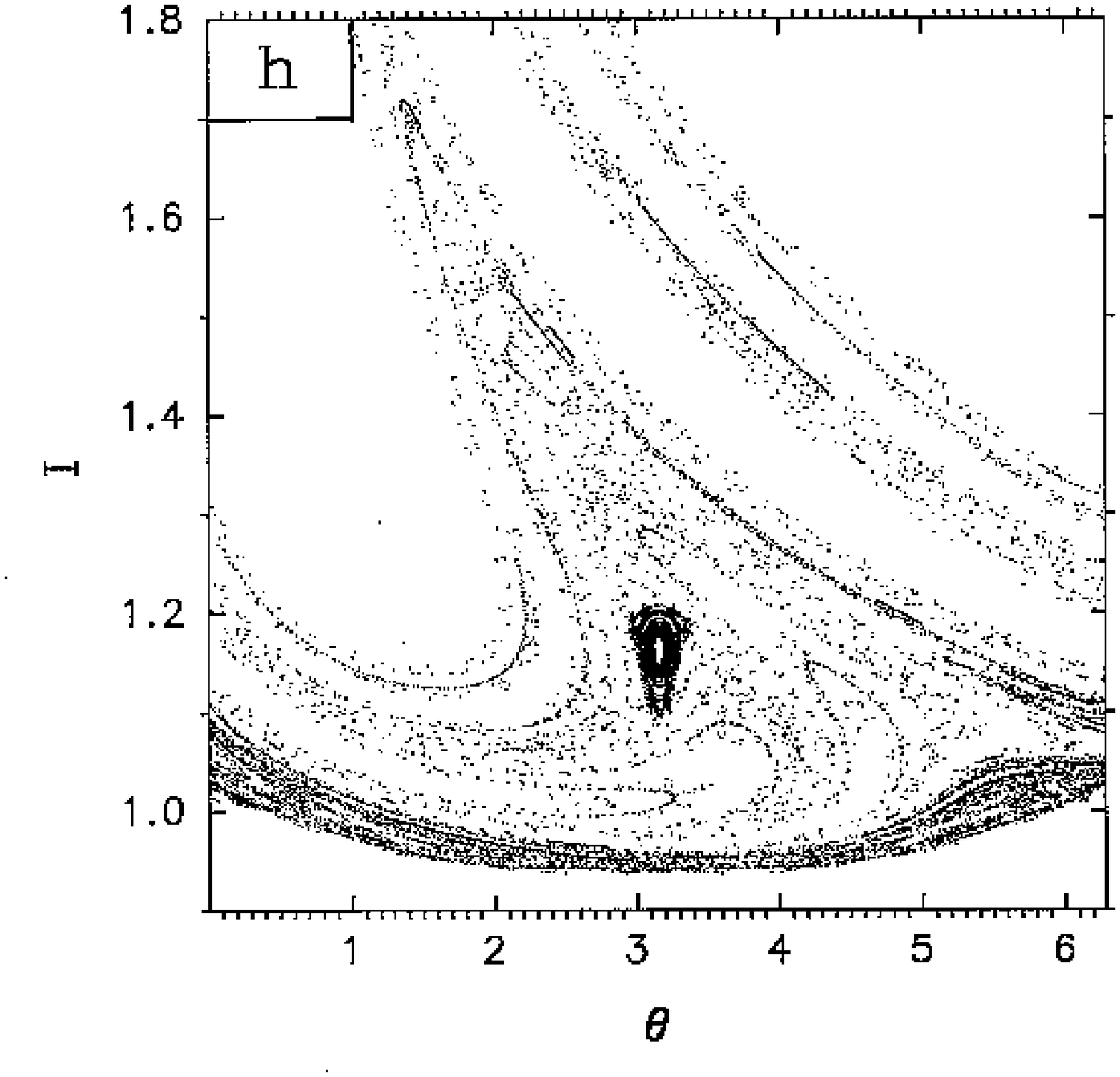,width=0.9\linewidth}
\end{figure}
\newpage
.

\begin{figure}[htbp]
\centering\epsfig{file=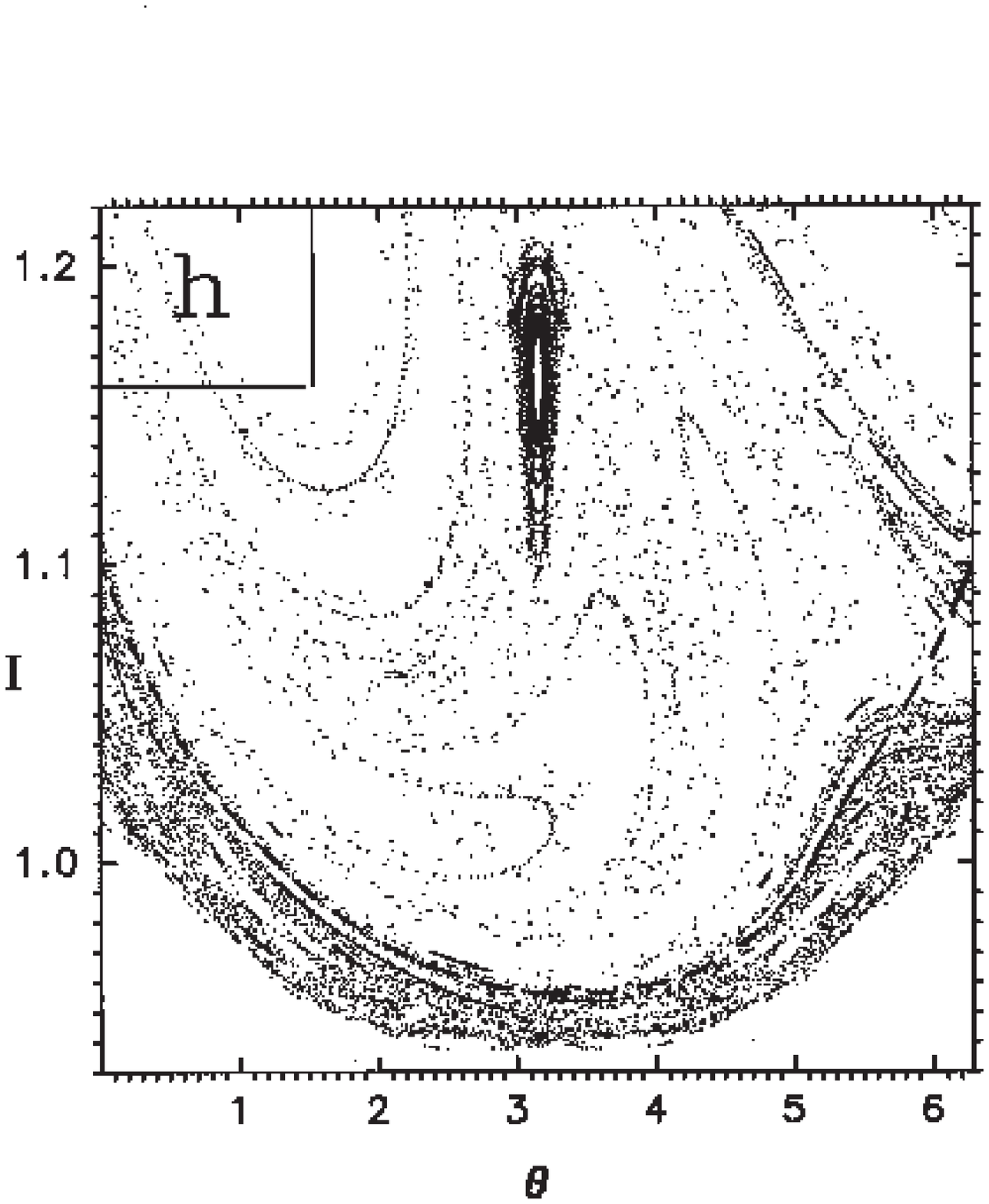,width=0.9\linewidth}
\caption{a) Surface of section (Poincar\'{e} map) for $\omega'_0 = 0.6876$,
$F_0 = 0.045$, $F_{0S} = 0.02777$ and $\phi_0 = \pi/2$. b) detail;
as in Fig. \ref{fig9}, the dashed line is an approximate separatrix
curve.}
\label{fig10}
\end{figure}
\newpage
.

\begin{figure}[htbp]
\centering\epsfig{file=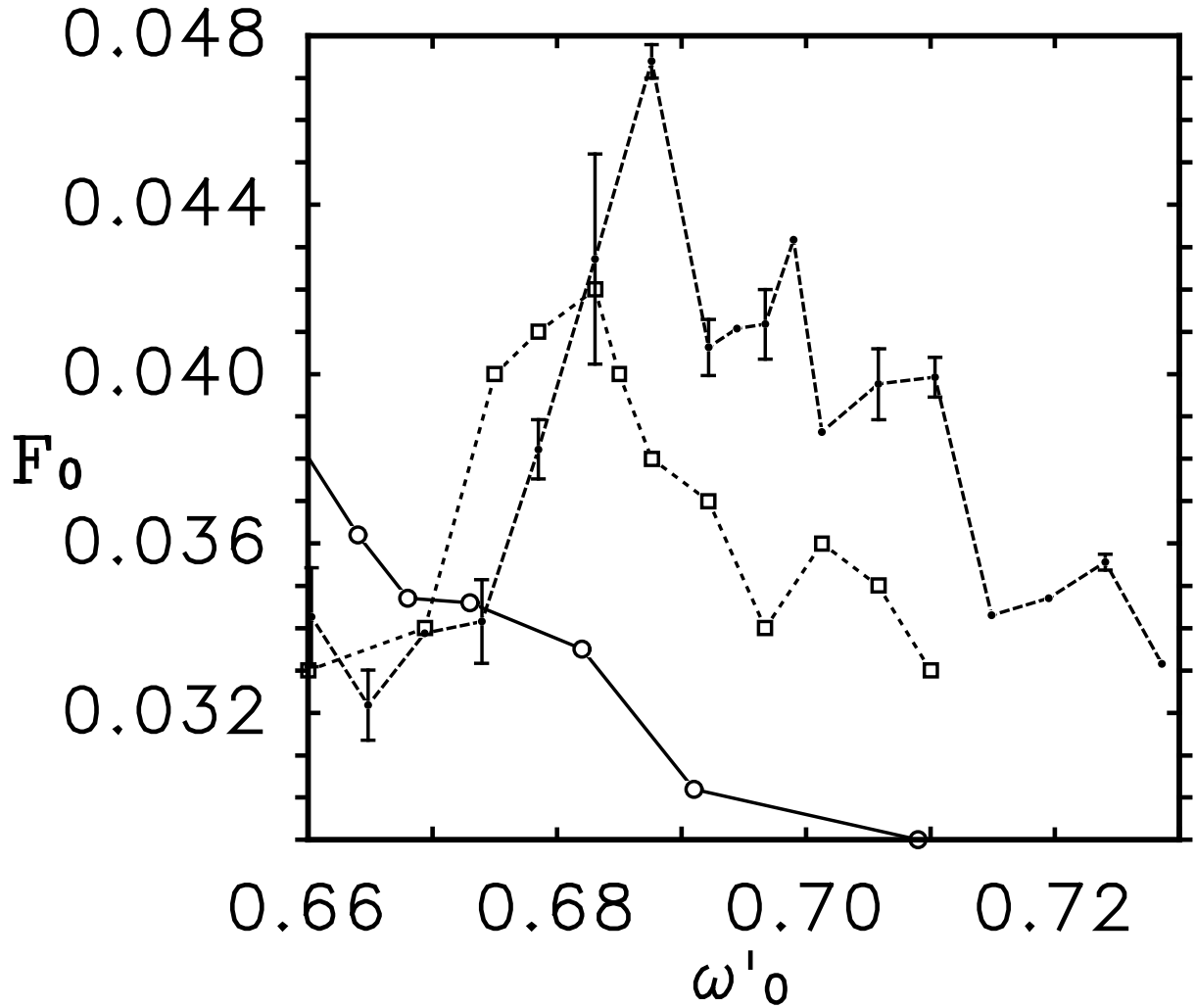,width=0.9\linewidth}
\caption{Rescaled peak microwave field for $10\%$ ionization probability $F_0(10\%)$ versus rescaled
frequency $\omega'_0$. Quantum results: squares. Experimental
results: dots. Classical results: circles. The classical curve
shows no sign of the quantum and experimental peak at
$\omega'_0\simeq 0.68$. (From Ref. \cite{ref40})}
\label{fig11}
\end{figure}
\newpage
.

\begin{figure}[htbp]
\centering\epsfig{file=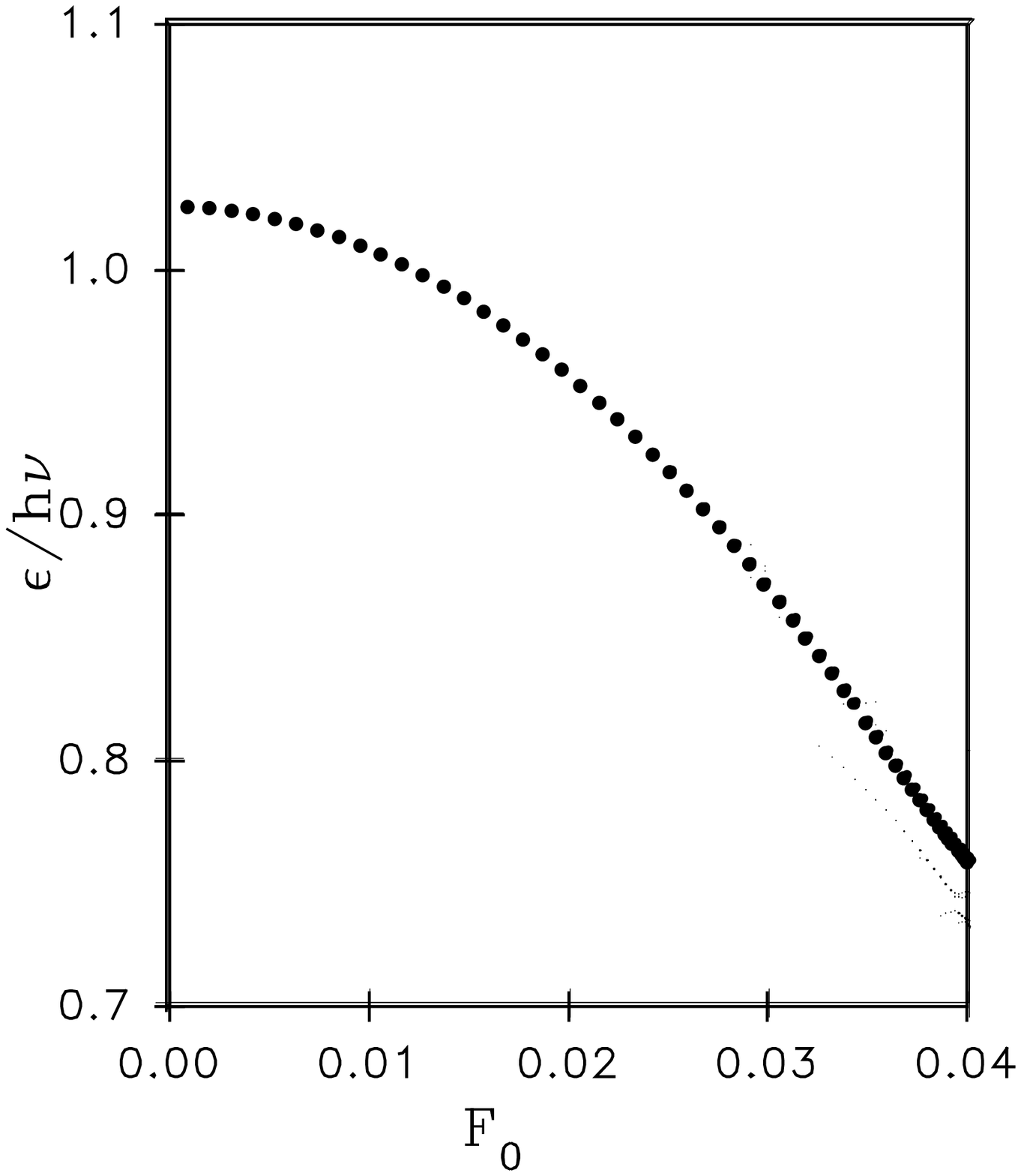,width=0.9\linewidth}
\end{figure}
\newpage
.

\begin{figure}[htbp]
\centering\epsfig{file=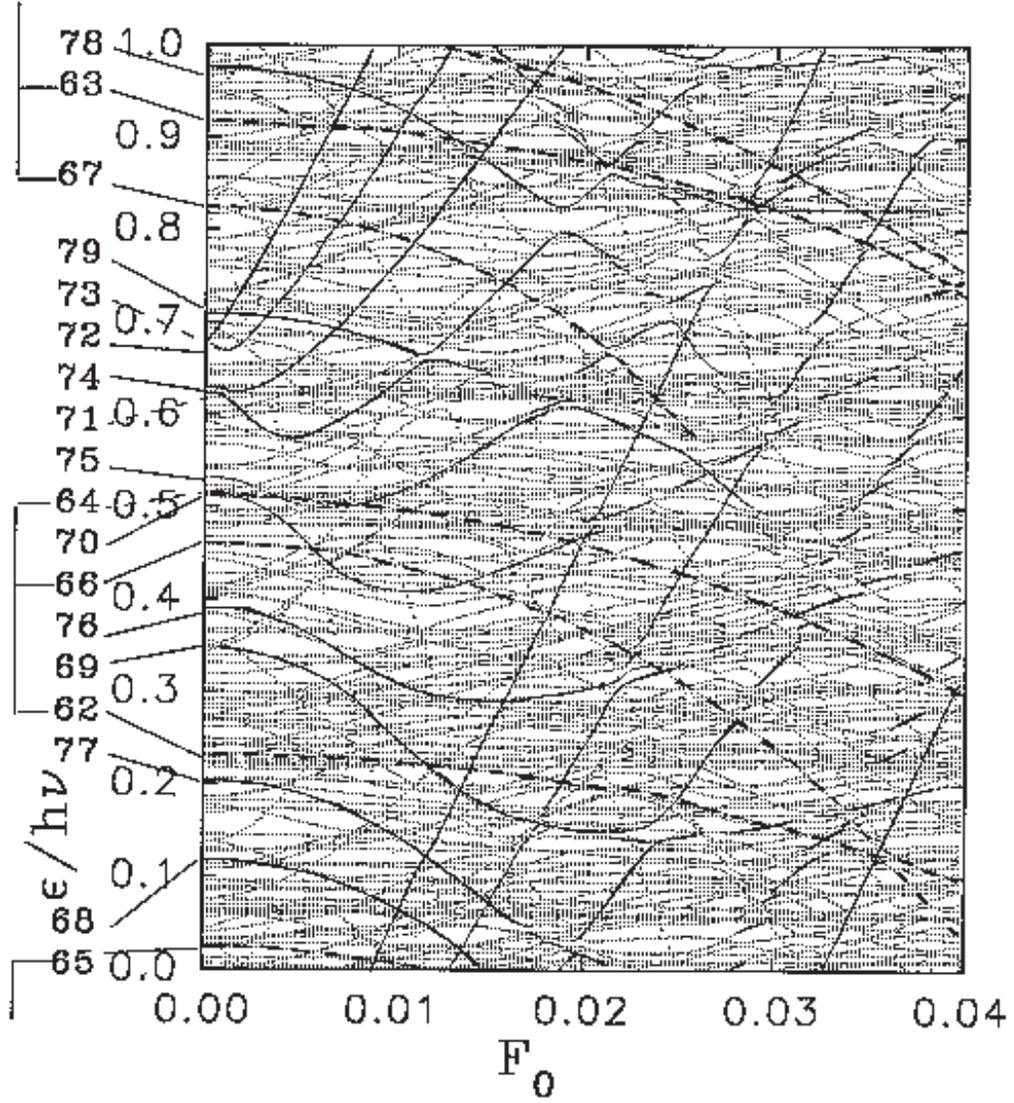,width=0.9\linewidth}
\caption{a) The evolution of the projections of the wavefunction on the instantaneous Floquet states at
every period of the rising edge of the microwave pulse. The
parameters are: $\omega'_0= 0.675$, $F_0^{max} = 0.04$, $F_{0S}
=0.02777$, $T = 116$ microwave periods and $n_0 = 65$. The three states making up most of the probability are
$n= n_0 = 65$ (full line), $n=63$ (dashed line) and a self ionizing
state (dotted line). b) Quasienergy curves for the rescaled (to
$n_0 = 65$) parameters $\omega'_0= 0.675$ and  $F_{0S}
=0.02777$; the horizontal scale is the rescaled microwave field strength. The
curves of the levels belonging to the $\omega'_0= 1/1$ nonlinear
quantum resonance are marked as full lines; the two groupings of
those belonging to the $\omega'_0= 2/3$ nonlinear quantum resonance
as dashed lines.}
\label{fig12}
\end{figure}
\newpage
.

\begin{figure}[htbp]
\centering\epsfig{file=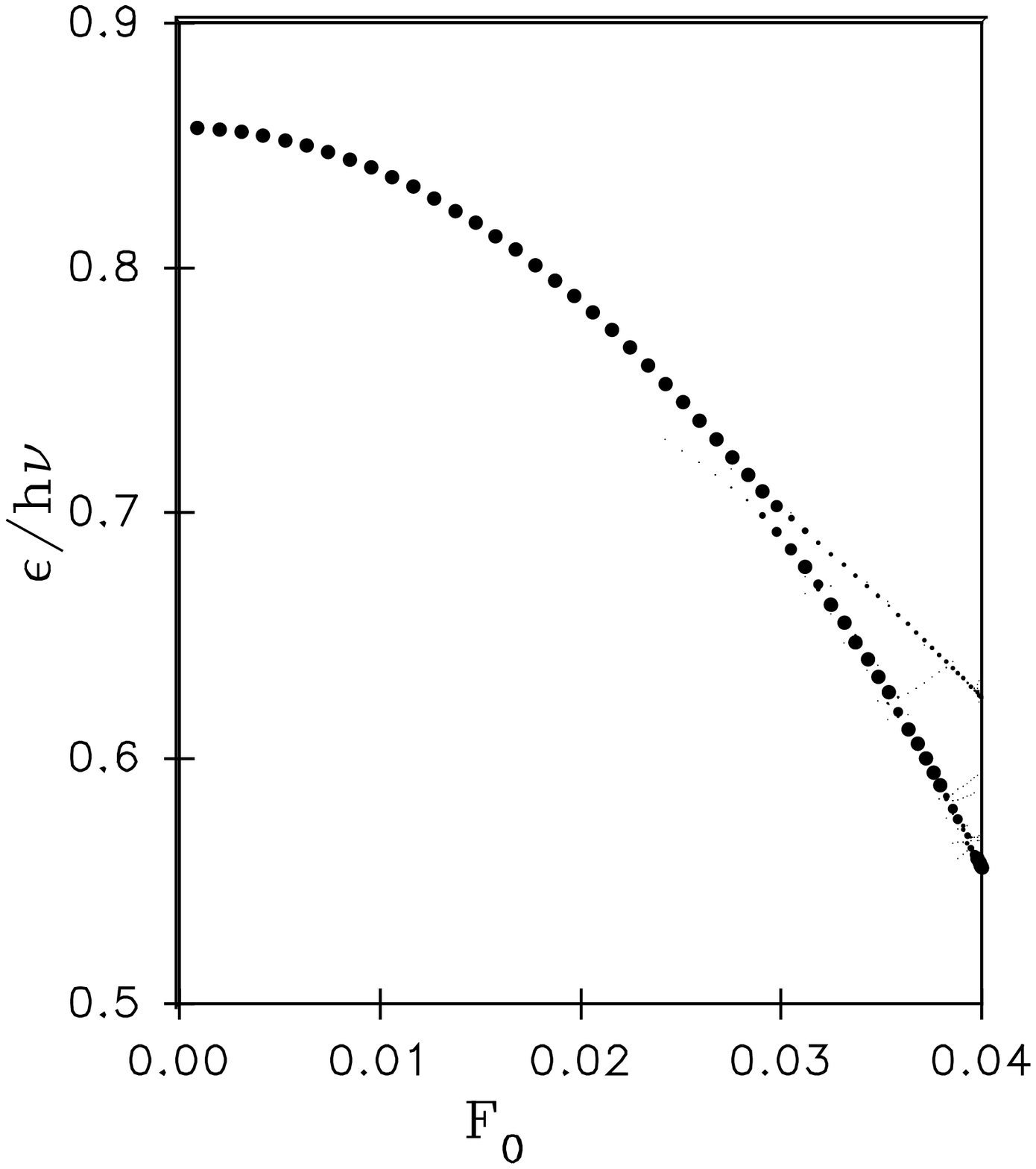,width=0.9\linewidth}
\end{figure}
\newpage
.

\begin{figure}[htbp]
\centering\epsfig{file=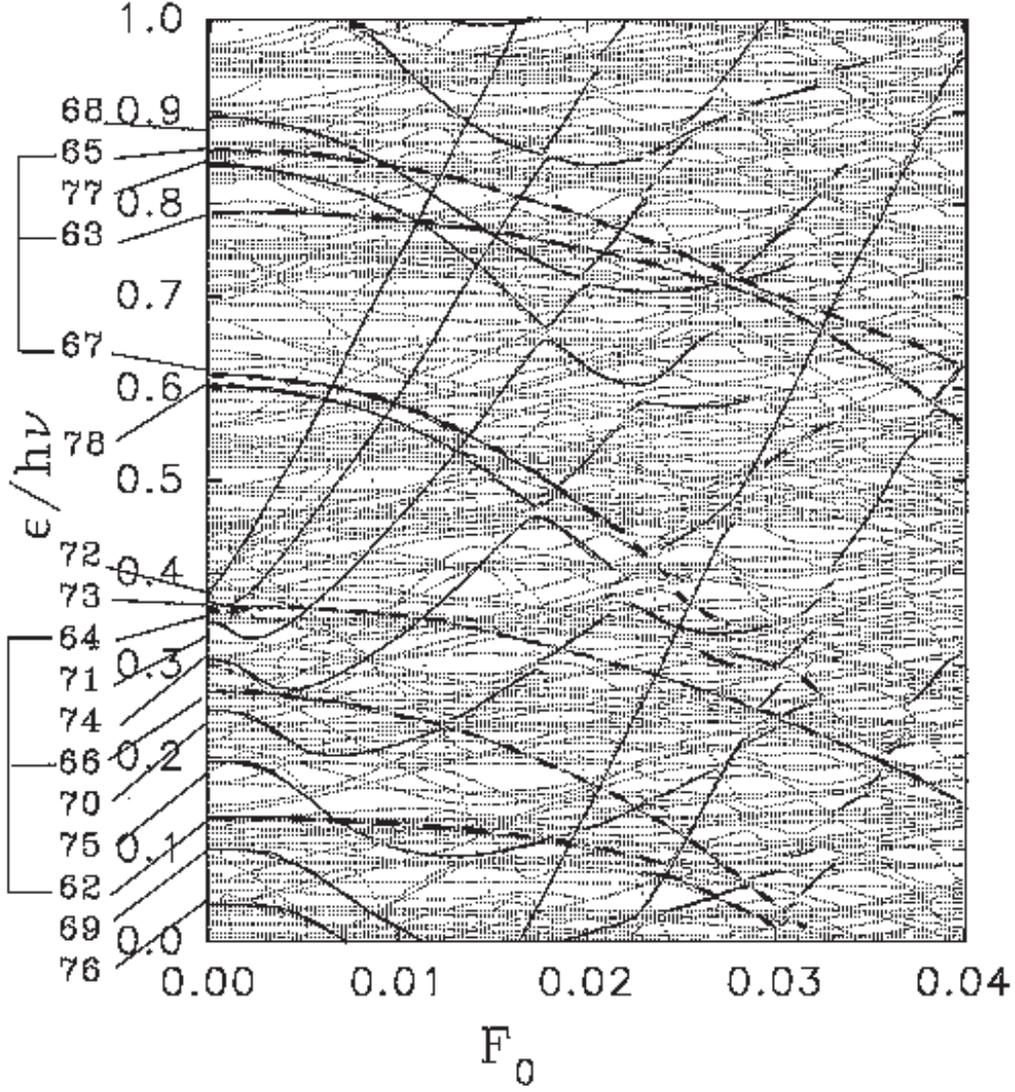,width=0.9\linewidth}
\caption{a) The evolution of the projections of the wavefunction on the instantaneous Floquet states at
every period of the rising edge of the microwave pulse. The
parameters are: $\omega'_0 = 0.685$, $F_0^{max} = 0.04$, $F_{0S}
=0.02777$, $T = 116$ microwave periods and $n_0 = 65$. The two states making up most of the probability are $n=
n_0 = 65$ (full line) and $n=63$ (dashed line). b) Quasienergy
curves for the rescaled (to $n_0 = 65$) parameters $\omega'_0 =
0.685$ and $F_{0S} =0.02777$; the horizontal scale is the rescaled
microwave field strength. The curves of the levels belonging to the
$\omega'_0= 1/1$ nonlinear quantum resonance are marked as full
lines; the two groupings of those belonging to the $\omega'_0= 2/3$
nonlinear quantum resonance as dashed lines.}
\label{fig13}
\end{figure}
\newpage
.

\begin{figure}[htbp]
\centering\epsfig{file=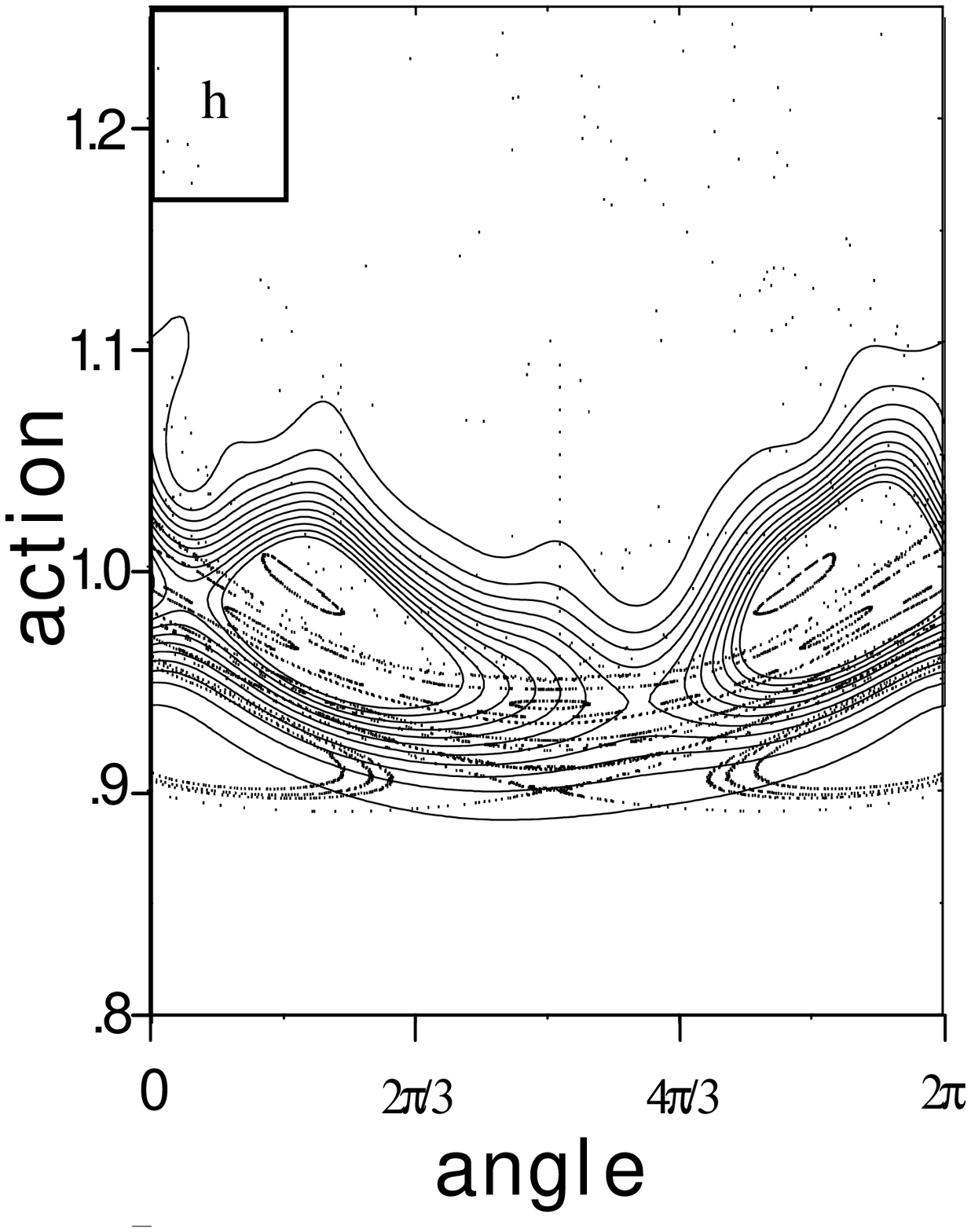,width=0.9\linewidth}
\end{figure}
\newpage
.

\begin{figure}[htbp]
\centering\epsfig{file=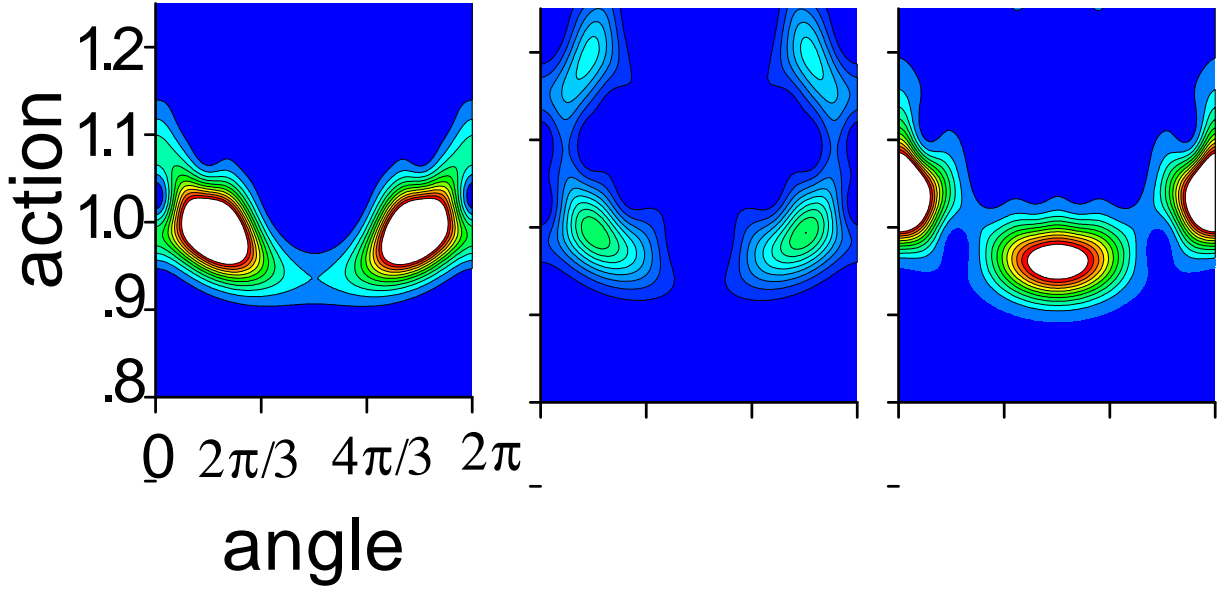,width=0.9\linewidth}
\caption{a) Husimi function of the quantum wavefunction at the peak of the pulse superimposed over
the classical surface of section at the same time. The parameters
are $\omega'_0 = 0.675$, $F_0^{max} = 0.04$, $F_{0S}
=0.02777$, $T = 116$ microwave periods and $n_0 = 65$. b) Husimi functions of the three eigenstates on which the
above instantaneous wavefunction has the highest projections: in
the order two states supported by the two islands of the $\omega'_0
=2/3$ classical resonance zone ($62.3\%$ and $15.7\%$ respectively) and a
state supported by the chaotic separatrix zone ($12.7\%$).}
\label{fig14}
\end{figure}
\newpage
.

\begin{figure}[htbp]
\centering\epsfig{file=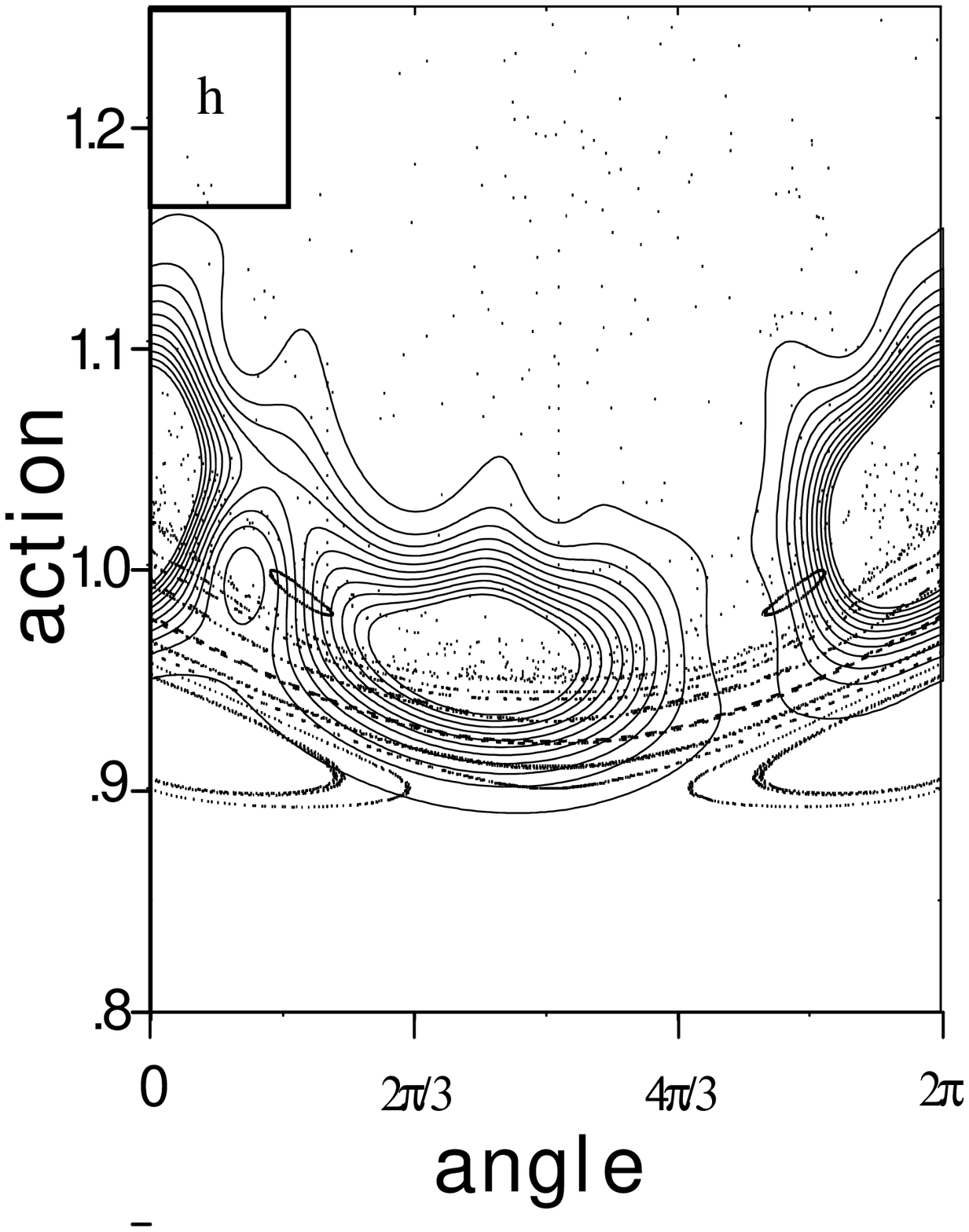,width=0.9\linewidth}
\end{figure}
\newpage
.

\begin{figure}[htbp]
\centering\epsfig{file=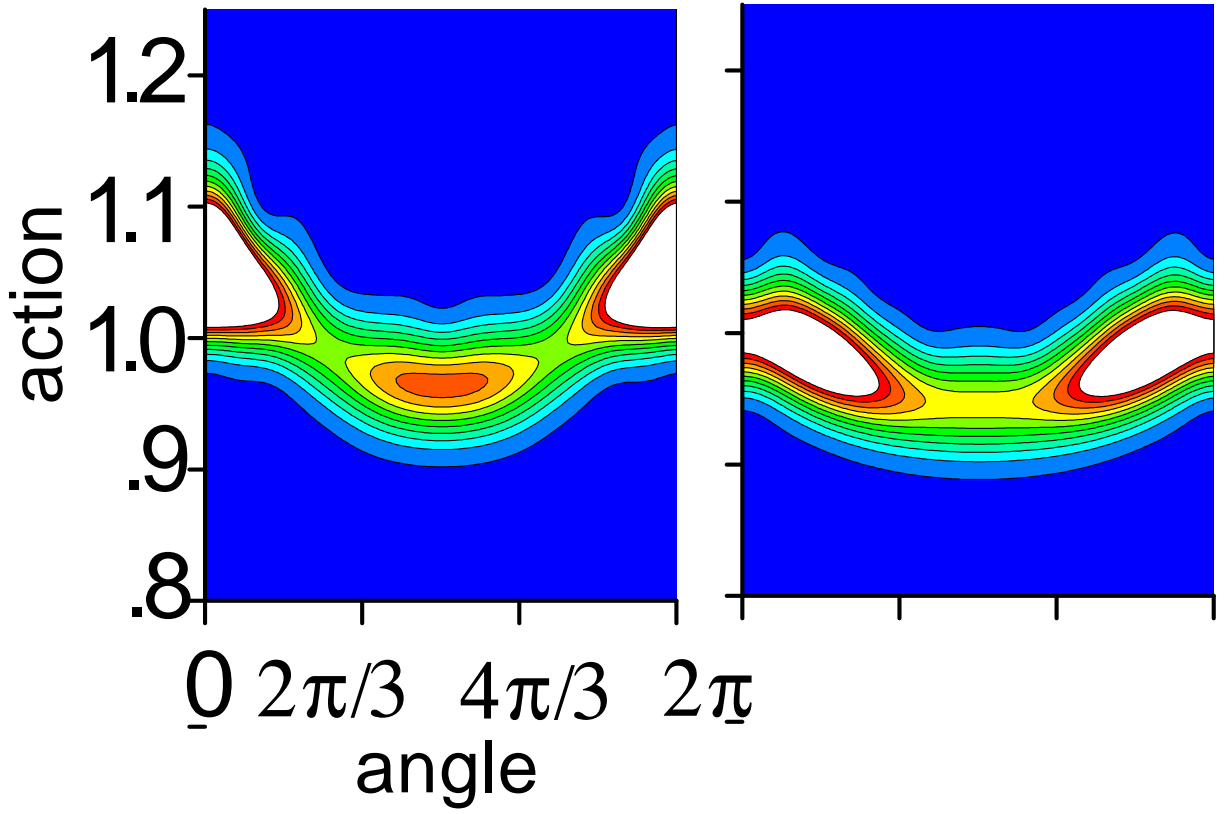,width=0.9\linewidth}
\caption{a) Husimi function of the quantum wavefunction at the peak of the pulse superimposed over
the classical surface of section at the same time. The parameters
are $\omega'_0 = 0.685$, $F_0^{max} = 0.04$, $F_{0S}
=0.02777$, $T = 116$ microwave periods and $n_0 = 65$.
b) Husimi functions of the two eigenstates on which the above
instantaneous wavefunction has the highest projections: in the
order a state mostly localized around the unstable fixed point of
the $\omega'_0 =1/1$ classical resonance zone ($61.6\%$) and a
state mostly supported by the two islands of the $\omega'_0  =2/3$
classical resonance zone ($19.3\%$).}
\label{fig15}
\end{figure}
\newpage
.

\begin{figure}[htbp]
\centering\epsfig{file=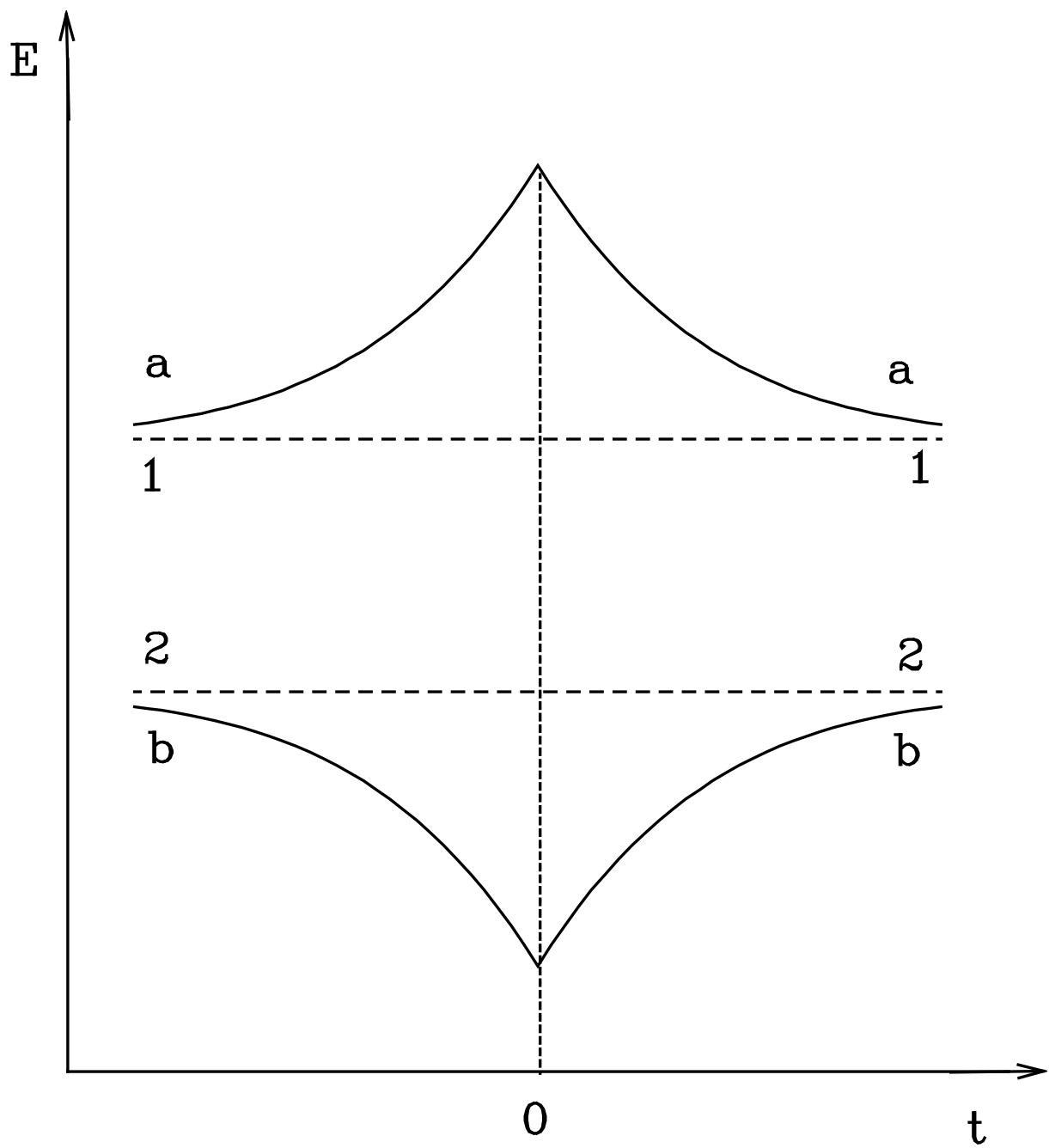,width=0.9\linewidth}
\caption{Behaviour of the adiabatic (full lines) and diabatic (dashed lines) levels in the Demkov
model as a function of time; the peak of the pulse is reached at
$t=0$. (From Ref. \cite{ref40})}
\label{figc1}
\end{figure}
\newpage
.

\begin{figure}[htbp]
\centering\epsfig{file=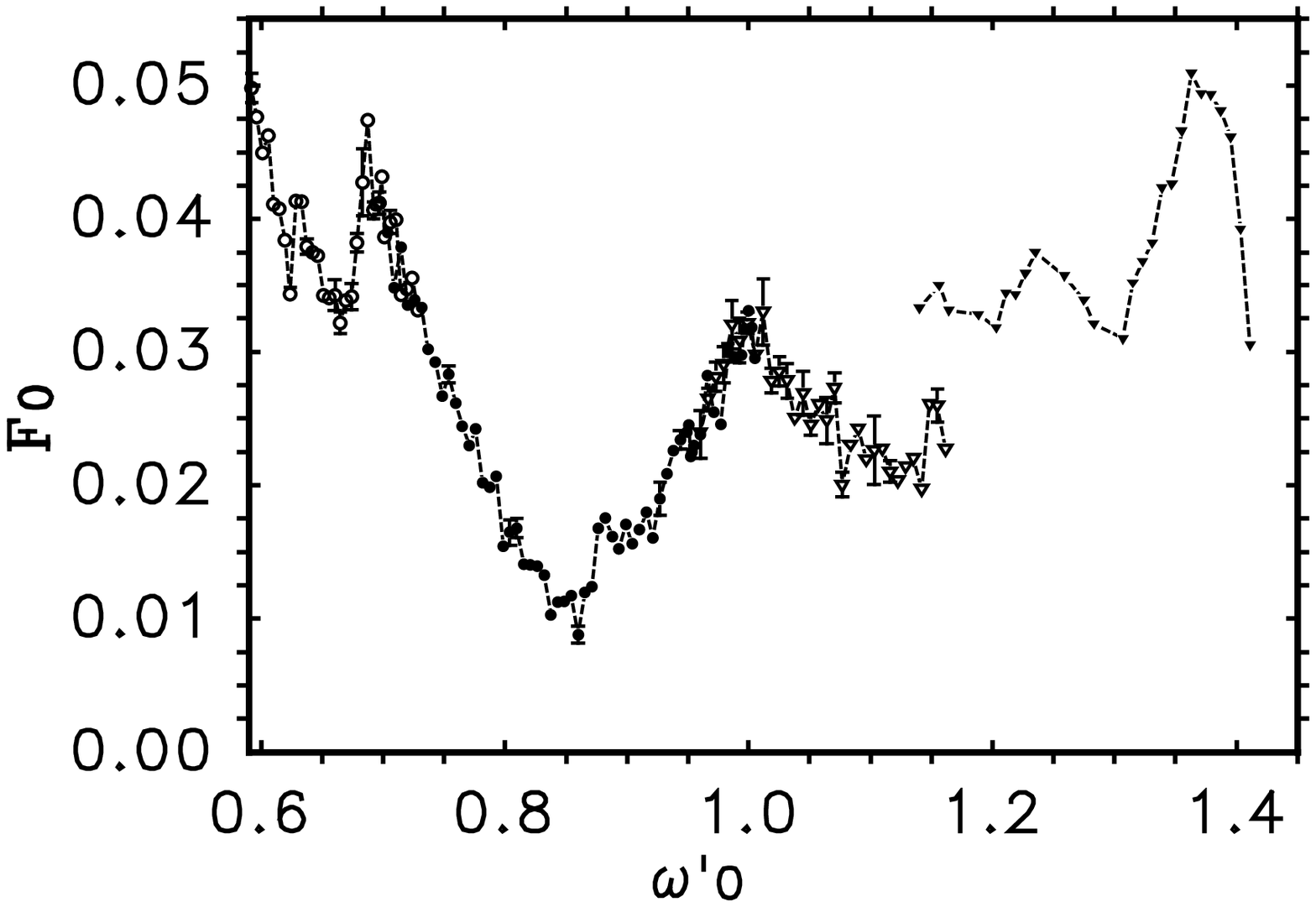,width=0.9\linewidth}
\end{figure}
\newpage
.

\begin{figure}[htbp]
\centering\epsfig{file=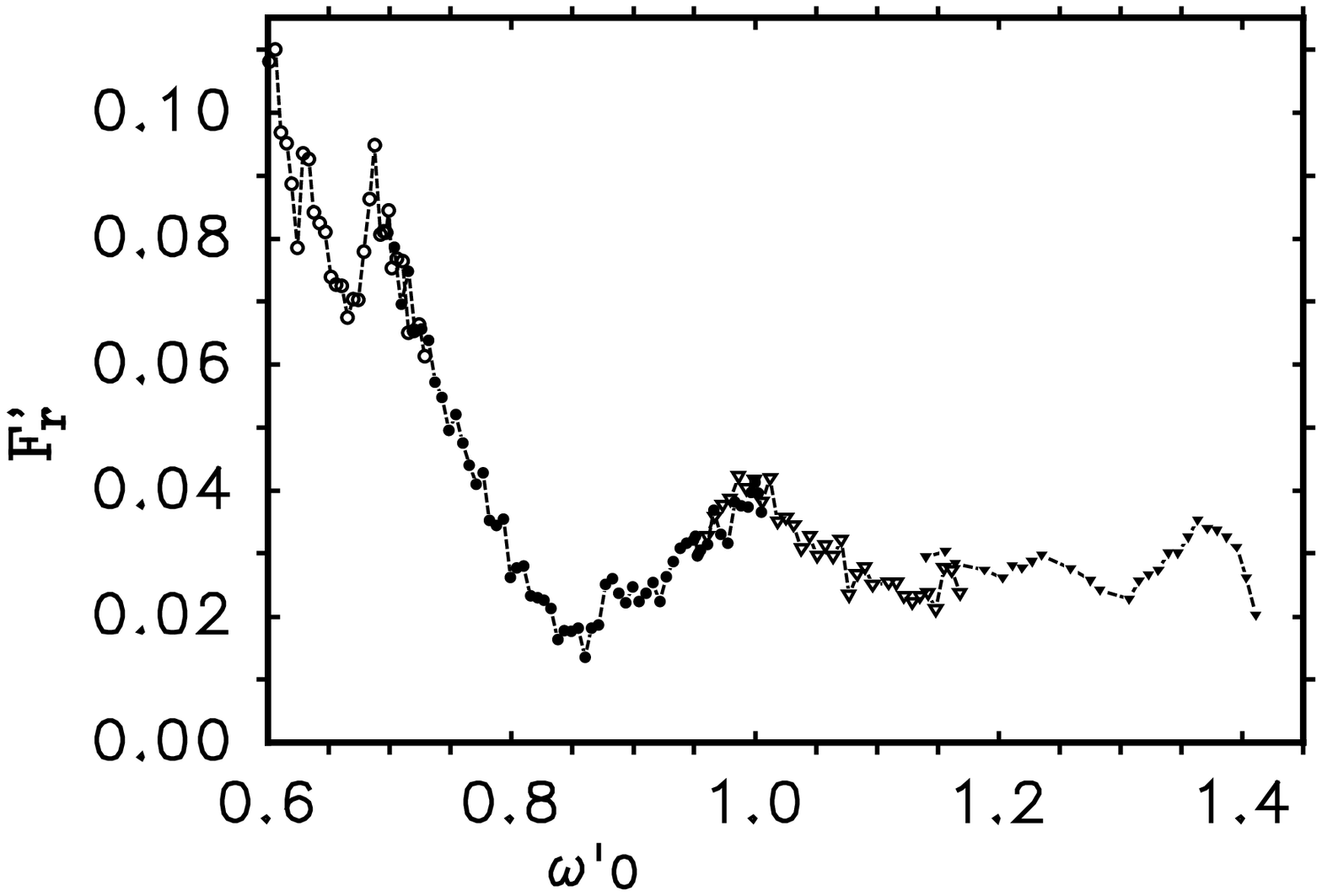,width=0.9\linewidth}
\caption{Peak microwave field for $10\%$ ionization probability a) rescaled to $n_0$ and b) rescaled to the
central action of the principal primary resonance and corrected for
the static field, both versus rescaled frequency $\omega'_0$:
experimental results from Pittsburgh. circles: $n_0= 65$, $F_S =
8\hspace{.1in} V/cm$; dots: $n_0= 69$, $F_S = 8\hspace{.1in} V/cm$;
empty triangles: $n_0= 72$, $F_S = 8\hspace{.1in} V/cm$; triangles:
$n_0= 80$, $F_S = 1\hspace{.1in} V/cm$. The sharp break between the
$F_S = 8\hspace{.1in} V/cm$ data and the $F_S = 1\hspace{.1in}
V/cm$ ones in a) has disappeared with the new choice of rescaling. (From Ref. \cite{ref40})}
\label{figd1}
\end{figure}


\begin{thebibliography}{47}
\bibitem{ref13}I have shown it analytically only in pendulum approximation
\cite{ref17} (see appendix A for a breaf description of it); but a
direct measurement of the distances between the (numerically
calculated) quasienergy levels of the principal primary quantum
nonlinear resonance shows that they are equally spaced also for the
full problem.
\bibitem{ref14}Similarly, in the same limit, states with $k \gtrsim 0$ are essentially
excited harmonic oscillator states forced to oscillate at the
microwave frequency (see ref. \cite{ref8}).
\bibitem{ref38}To find $n_s$ we first calculate the resonant action $I_r$; $n_s$ is
then the integer closest to $2I_r- n_0$. Since $I_r$ is calculated
making use of the approximate formula $\omega I_r^3/ (1- 3F_S
I_r^4) = 1$, in same cases the $n_s$ thus calculated can be off by
one; we therefore always numerically calculate the zero microwave
field quasienergies for $n_s$, $n_s+1$ and $n_s-1$; the real $n_s$
is the one of these whose quasienergy is closest to that of $n_0$.
\bibitem{ref39}In particular, in the cases when other peaks are present, the
quantum final population on the peak centered at $n_0$ displays
oscillations as a function of $F^{max}$ that are absent from the
corresponding classical simulations. The absence of classical
oscillations is due to the scrambling of the phases induced by the
chaotic band (using the integrable pendulum approximation and
parameters comparable to the present ones, oscillations are present
\cite{ref34}); their presence in the quantum simulation seems to
indicate that the quantum system is at least partially blind to the
chaotic band at the separatrix. The scarring itself of the
separatrix FE can be interpreted as expression of this blindness
\cite{ref52}.
\bibitem{ref51}A similar indifference of the quantum system to chaotic bands
and agreement with the behaviour expected from just the resonant
term has been observed in other models: see e.g. ref. \cite{ref52}.
\bibitem{ref41}The initial decay of about $80\%$ in the population of this state
and of about $20\%$ of the $n=65$ state it is ``crossing" happens
in less than one microwave period; afterwards both states decay at
the same rate.
\bibitem{ref47}Actually, for an orbit within the resonance zone, both $\left<x \right>$ and
$\left<\varphi \right>$ are zero in pendulum approximation. This
means that the average position of the orbit is at the center of
the resonance zone, that is: it coincides with the point
representing in sectioned phase space the stable periodic orbit.
Quantization of the tori within the resonance zone will therefore
result in eigenfunctions centered on that same orbit and
oscillating in phase with it.
\bibitem{ref49}A simple example is in this case the pendulum Hamiltonian  $H=
H_0+V$ where $V=2V_0\lambda cos(\varphi)$, $H_0 =
A[I-(|\Delta|/A-1)/2)^2$ and we assume $\lambda
= e^{-B^2|t|}$. If we take $|\Delta|\approx 2A$ and we perform the usual
``semiclassical" quantization $I= -  i \partial /\partial \varphi$
on the base $|m> = e^{im\varphi}/\sqrt{2\pi}$, the two states $|0>$
and $|1>$ will be closer to each other than to any other state
interacting with them. It will therefore be possible (as long as
$V_0 \ll 2A$) to treat them as a two-level system with a
Hamiltonian matrix as the one given above.
\end{thebibliography}
\end{document}